\documentclass[blind,review,3p]{elsarticle}

\usepackage{lineno,hyperref}
\modulolinenumbers[5]

\usepackage{amssymb}

\usepackage{mathtools}
\usepackage{setspace}
\usepackage{longtable}
\usepackage{threeparttable}
\usepackage{multicol}
\usepackage{multirow}
\usepackage{pdflscape}

\usepackage{tikz}
	\usetikzlibrary{arrows,chains,matrix,positioning,scopes}
	\tikzset{>=triangle 45} 
\usepackage{pgfplots}
\usepackage{graphics}
	\graphicspath{{figures/}}
\usepackage{amsmath}
\usepackage{amssymb}
\usepackage{float}
\usepackage{multirow}
\usepackage{fourier} 
\usepackage{array}
\usepackage{makecell}

\usepackage{bm}

\usepackage{tikz}
\usetikzlibrary{shapes,arrows}
\tikzstyle{block} = [rectangle, draw,
    text width=5em, text centered, rounded corners, minimum height=4em]
\tikzstyle{line} = [draw, -latex']    
\tikzstyle{optional}=[dashed,fill=gray!50]
\tikzstyle{Text} = [ellipse, fill = white, draw = white, text = black, align = center, minimum height = 1cm, minimum width = 1cm]
\tikzstyle{VineNode} = [ellipse, fill = white, draw = black, text = black, align = center, minimum height = 1cm, minimum width = 1cm]
\tikzstyle{DummyNode}  = [draw = none, fill = none, text = black] 
\tikzstyle{TreeLabels} = [draw = none, fill = none, text = black] 
\newcommand{\xshiftNodes}{0.7*\linewidth}
\newcommand{\yshiftLabels}{-.25cm}  
 
\journal{arXiv}

\biboptions{authoryear}

\begin{document}

\begin{frontmatter}



\author[1]{Karoline Bax} 
\author[2]{Özge Sahin}
\author[2]{Claudia Czado}
\author[1]{Sandra Paterlini \corref{cor1}}


\address[1]{Department of Economics and Management, University of Trento, Trento, Italy}
\address[2]{Department of Mathematics, Technical University of Munich, Munich, Germany}

\title{ESG, Risk, and (tail) dependence}


\begin{abstract}

While environmental, social, and governance (ESG) trading activity has been a distinctive feature of financial markets, the debate if ESG scores can also convey information regarding a company's riskiness remains open. Regulatory authorities, such as the European Banking Authority (EBA), have acknowledged that ESG factors can contribute to risk. Therefore, it is important to model such risks and quantify what part of a company's riskiness can be attributed to the ESG scores. This paper aims to question whether ESG scores can be used to provide information on (tail) riskiness.  By analyzing the (tail) dependence structure of companies with a range of ESG scores, that is within an ESG rating class, using high-dimensional vine copula modelling,  we are able to show that risk can also depend on and be directly associated with a specific ESG rating class.  Empirical findings on real-world data show positive not negligible ESG risks determined by ESG scores, especially during the 2008 crisis.\\
\end{abstract}


\end{frontmatter}


\section{Introduction}

After the 2007-2009 financial crisis, many models which used to capture the dependence between a large number of financial assets were revealed as being inadequate during crisis. Moreover, research has shown that the dependence structure of global financial markets has grown in importance in areas of optimal asset allocation, multivariate asset pricing, and portfolio tail risk measures \citep{ref:xu}. Consequently, the enormous losses and the increased volatility in the global financial market elicited calls for an even more diligent risk management. 

Over the past decade,  the interest in socially responsible investments has grown exponentially \citep{ref:auer}.  The availability of non-financial data, including corporate social responsibility (CSR) or environmental, social, and governance (ESG) data, has skyrocketed and gained interest from investors for various reasons.  According to \cite{Li2020},  in 2019,  70 different firms were identified as providers of some sort of ESG rating. While some studies are pointing out ambiguity and divergence between these different ratings (see \cite{Berg2019, Berg2021, Billio2021,Gibson2020,Serafeim2021b}, the overall idea of ESG scores is to indicate a level of ESG performances. These scores are based on several criteria and measurements and are given by a rating institution \cite[]{ref:bhatta}. The rating institutions use quantitative and qualitative methods to assign an ESG score to a company \citep{ref:berg}.  In brief, companies are awarded large scores for ESG responsible behavior but are awarded low scores for ESG irresponsible behavior. Typically a company is associated with a rating class (i.e., $A,B,C$, or $D$) based on its ESG score value using thresholds or quartiles. Then, companies with the same ESG score or ESG scores within the same threshold or quartile are allocated in the same ESG rating class.  Acting as complementary non-financial information, ESG scores can have the potential to increase the accuracy in performance forecasts and risk assessments \citep{ref:achim}. 

In 2018, the European Commission published its \textit{Action Plan on Financing Sustainable Growth} which provided the European Union (EU) with a roadmap on sustainable finance and for future work across financial systems \citep{EBA2018}.  Furthermore, the increase in demand in socially responsible investments stems from investors and asset managers pressured by stakeholders to push companies to behave responsibly and improve their ESG strategy \citep{ref:henriksson}. The biennial 2018 Global Sustainable Investment Review states that over \$30 trillion had been invested with explicit ESG goals \citep{GSIA2018}.  According to a 2018 global survey,  more than 50\% of the international asset owners are currently considering or already implementing ESG scores in their investment strategy \citep{Consolandi2020}. Moreover,  \cite{Eccles2019} state that this interest in ESG assets is driven by the growing evidence of the positive impact of ESG materiality on financial performance.  Also, the European Banking Authority (EBA) stressed the importance of including ESG information into the regulatory and supervisory framework of EU credit institutions \citep{EBA2018}.  Furthermore,  there is a need to improve the measurement and modelling of the impacts of climate change on financial stability in order to underpin a policy debate as highlighted by the \cite{ECB2021}.  Nevertheless, using ESG factors has made the investment process noticeable more complex \citep{ref:berg}.  
While more than 2000 empirical studies have been done analysing ESG factors and financial performance,  little is known about the dependence structure and associated risks \citep{ref:friede, ref:shafer, new:loof2021}.  This is especially important as ESG scores are often linked to investment risk.  The EBA defines ESG risks as “the risks of any negative financial impact to the institution stemming, from the current or prospective impacts of ESG factors on its counterparties" \citep[p. 28]{EBADiscussion}.  While regulators finally acknowledge the role of ESG factors in determining part of a company's risk, no guidelines are provided in how to best capture and quantify such risk.  Then,  ESG scores should possibly deliver some information on a company's ESG risk and its riskiness as a whole.  This also means that companies that have the same ESG rating, as explained above,  should share similar risk characteristics and properties. Understanding these components is especially important in times of increased volatility as considering the (tail) dependence structure among similarly rated assets is essential in order to set in place effective risk management and diversification strategies \citep{ref:ane, ref:frahm, ref:malevergne, ref:berg}. 

The contribution of this research is manifold. We propose a R-vine copula ESG risk model to capture the dependence using vine copulas \citep{ref:bedford2, ref:bedford1, ref:aas,  ref:joe2014,ref:czado1,CzadoNagler2022} and to identify the systematic and the idiosyncratic risk component considering specific ESG classes of each asset. Inspiration for this model arise from \cite{ref:brechmann} who considered sectorial dependencies. This allows us to show, to our knowledge for the first time, that risk can also be linked to an ESG rating classes.  Furthermore, it contributes to the understanding of the (tail) dependence structure of various assets belonging to the same ESG rating class and introduces different risk measures that try to capture specific ESG risk and the market risk conditionally on ESG classes. 
By quantifying the overall and lower tail ESG risk among assets that belong to the same ESG class,  we show that these dependencies exist, can be quantified, and are not negligible, especially in times of crisis.

The paper is structured as follows; Section 2 summarizes the literature and creates an understanding of ESG scores and the occurrence of (tail) risk and dependence structures while Section 3 describes the S\&P 500 data used and includes a preliminary risk analysis.  Section 4 introduces dependence modelling and vine copulas and then proposes the R-vine copula ESG risk model.  Section 5 reports the empirical results.  Lastly,  Section 6 concludes and provides an outlook for future research.

\section{ESG Scores,  Dependence, and Risk}
While ESG scores solely try to capture the amount of positive ESG disclosure  of a company, the large influx into ESG investments have brought attention to the risk and return from such investments.
 It has been shown that ESG factors may impact financial performance by substantiating themselves in “financial or non-financial prudential risks, such as credit, market, operational, liquidity and funding risks" \citep[p. 27]{EBADiscussion}. According to the EBA,  ESG risks are defined to materialize when ESG factors have a negative impact on the financial performance or solvency \citep{EBADiscussion}. Furthermore, it is argued that the materiality of ESG risks depends on the risks posed by ESG factors over different time frames \citep{EBADiscussion}. If this is accurate, the ESG scores and ratings should contain information about the company's risk. Even though ESG has mostly been defined in terms of risk by the regulator, and there is an ongoing debate on the effects of using ESG scores on the financial performance research, there is no consensus on the (tail) dependence structure of assets within and between each ESG rating class. Understanding the (tail) dependence and risk structure of several assets, however, is necessary to access inherent risks in the financial market. 

Generally, market returns are assumed to follow a multivariate normal distribution; however, research has shown that this is  found to be not accurate in reality,  and left tails are often heavier than right tails \citep{Cont2001, ref:jondeau}. Therefore, modelling the possibly non-Gaussian dependence among assets and  understanding the appearance of joint (tail) risk is especially important for asset pricing and risk management.  Tail risk or tail dependence is characterized as the probability of an extremely large negative (positive) return of an asset given that the other asset yields an extremely large negative (positive) return and is commonly quantified by the so-called tail-dependence coefficient \citep{Embrecht2001, ref:frahm,  ref:xu}. Tail risk arises when the likelihood of an extreme event that is more than three standard deviations away from the mean is more likely to occur than shown by a normal distribution \citep{ref:kelly}.  Generally, it is accepted that tail dependence can be used as a proxy of systemic risk, and tail risk has been linked to negative consequences for corporate investment and risk-taking \citep{ref:gormley2, ref:gormley1, ref:shir}.

Some researchers argue that responsible ESG practices might mitigate the market's perception of a company's tail risk and, therefore, reduce ex-ante expectations of a left-tail event \citep{ref:shafer}. In brief, they state that considering responsible business practices when creating an equity portfolio can act as insurance against left-tail risk (defined as stock price tail risk using the slope of implied volatility) and, with that, protect company value.  This is also in line with \cite{De2015} as well as  \cite{ref:wamba} who add that especially positive environmental performance can act as insurance for companies, reducing the probability of an adverse event occurring and with that reducing the company's systematic risk. \cite{Kumar2016} add that positive ESG practices can make a company "less vulnerable to reputation, political and regulatory risk and thus leading to lower volatility of cash flows and profitability" (p. 292).  Furthermore, positive ESG performance may generate more loyalty from customers and employees and, through that, protect companies from unforeseen harmful events, resulting in reduced tail risk \citep{ref:shafer}. Besides, better ESG performance allows companies to experience adverse events less often and lose less value if they do occur \citep{ref:minor}.  On the other hand, \cite{Zhang2021} extend the work by \cite{ref:shafer} by considering implied skewness and find a higher negative tail risk for higher ESG rated companies.

Others recognize that improving ESG performance can also help with risk control and exposure \citep{ref:giese}. \cite{Maiti2020} finds that overall portfolios developed using the overall ESG as well as the individual E (Environment), S (Social), and G (Governance) factors generally show better investment performance, implying policy modifications. In contrast, \cite{ref:breedt} show that ESG-tilted portfolios do not necessarily have higher risk-adjusted returns.  These findings are in line with \cite{new:loof2021} who argue, focusing on the Covid-19 crisis, that higher ESG scores are associated with less downside tail risk but also lower upside potential on returns.  Moreover, \cite{Sherwood2017} find that integrating ESG information into the emerging market equity investment of institutional investors can create higher returns and lower downside risks compared to non-ESG equity investments.  Other scholars have observed a mitigating effect of ESG performance on stock price crash risk if a company has less effective governance \citep{ref:kim}.  \cite{ref:hoepner} found supporting evidence on one single institutional investor that ESG engagements can be associated with subsequent reductions in downside risk.  

Looking at a country's creditworthiness, \cite{Capelle2019} show that countries with above-average ESG scores are linked to reduced default risk, and smaller sovereign bond yield spreads.  Additionally,  \cite{Breuer2018} state that the cost of equity is reduced when a company invests in CSR, given that it is located in a country with high investor protection.  Moreover, \cite{ref:li} find that companies in regions of high social trust tend to have lower tail risk.  A possible reason why environmental practices have the power to control tail risk could be that a company's good social records will be more valuable in the long run due to a lower frequency of litigation \citep{ref:goldreyer}. Additionally, higher environmental standards are valuable to shareholders due to companies avoiding litigation costs, reputation losses, and environmental hazards \citep{ref:chan}. Furthermore,  ESG can drive an asymmetric return pattern in which socially responsible investment (SRI) funds (using positive rather than negative screenings) outperform conventional funds in times of crisis but underperform in calm periods \citep{ref:nofsinger}.  Moreover, \cite{Bae2019} find that CSR reduces the costs of high leverage and decreases losses in market share when firms are highly leveraged.  

Despite these extensive efforts, the effect of ESG scores on (tail) dependence and (tail) risks has not yet been clearly understood, and literature studying the impact of ESG behaviors on tail risk is still very limited \citep{Zhang2021}. While some question whether an ESG‐related risk factor that can help to identify superior investments even exists \citep{Cornell2021}, others provide evidence that ESG ratings are subject to a non-diversifiable risk component as the development depends on the overall market \citep{Dorfleitner2016}.  Therefore, in order to estimate ESG risk, a closer look the assets within the ESG class and possibly the complete market has to be taken. Investors have been aware of the need to use more sophisticated models to assess the dependence behavior of assets since the financial crisis, however, such models,  which allow investors to quantify (tail) dependence and (tail) risk among ESG-based classes of assets,  have yet to be introduced. 
In this paper, we apply vine copulas  to model the complex dependence structure and compute different risk measures, including ESG risk, market risk conditionally on ESG class, and an idiosyncratic risk component, which allow to explicitly capture (tail) dependence and possibly aid the investor in their decision-making process.

\section{The Data}\label{data}
We consider daily logarithmic return and yearly environmental, social, and governance (ESG) data of 334 US companies $j$, constituents of the  S\&P 500 index, for which the yearly ESG scores ($ESG_{y,j}$) from Refinitiv \footnote{The ESG scores considered are the ones available from Refinitiv,  the financial and risk business unit of Thomas Reuters.  Refinitiv offers a comprehensive ESG database across more than 500 different ESG metrics \citep{Refinitiv2021}. These percentile-ranked ESG scores are designed to measure the company’s relative ESG performance, commitment, and effectiveness across ten main topics, including emissions,resource use,  and human rights, all of which are based on publicly reported data \citep{Refinitiv2021}. }, which are ranked based and have values between 0 and 100,  are available in the period from 3 January 2006 to 31 December 2018. 
\cite{Refinitiv2021} allows to attribute the assets into four different rating classes based on their ESG score. Assets are given a D grade for an ESG score lower than 25, C for an ESG score between 25 and lower than 50, B for an ESG score between 50 and lower than 75, and finally A for an ESG score between 75 and 100.
The time period is chosen in order to have the largest possible time period for the largest number of assets to get a comprehensive analysis of assets belonging to the S\&P 500. The overall time frame is split into three different time periods $q$: $q_1= 2006-2010, q_2= 2011-2015$,  and $q_3= 2016-2018$.  The first two intervals are made of five years of data each, where the first interval includes the 2008 financial crisis. The last interval consists of three years of data, and their ESG scores are to be considered not definitive according to Refinitiv  as they were not yet five years old when downloading the data (in November 2020).  This means that the ESG scores can be modified from the provider post-publication. It is, however, not an option for investors to wait until scores are definitive as investment decisions are made daily using all the data available at the time.  
 In order to compute the weighted ESG class indices, which we later need in our R-vine copula ESG risk model, the  S\&P 500 market capitalization weights ($M_{j}$) from 1 January 2015, as well as the information on the economic sector $S$ for each asset $j$ are downloaded from \cite{Refinitiv2021}.

\subsection*{95\% VaR}
	
\begin{figure}[H]
    \centering
    \begin{minipage}{.5\textwidth}
        \centering
        \vspace{0.75cm}
\includegraphics[scale=0.38]{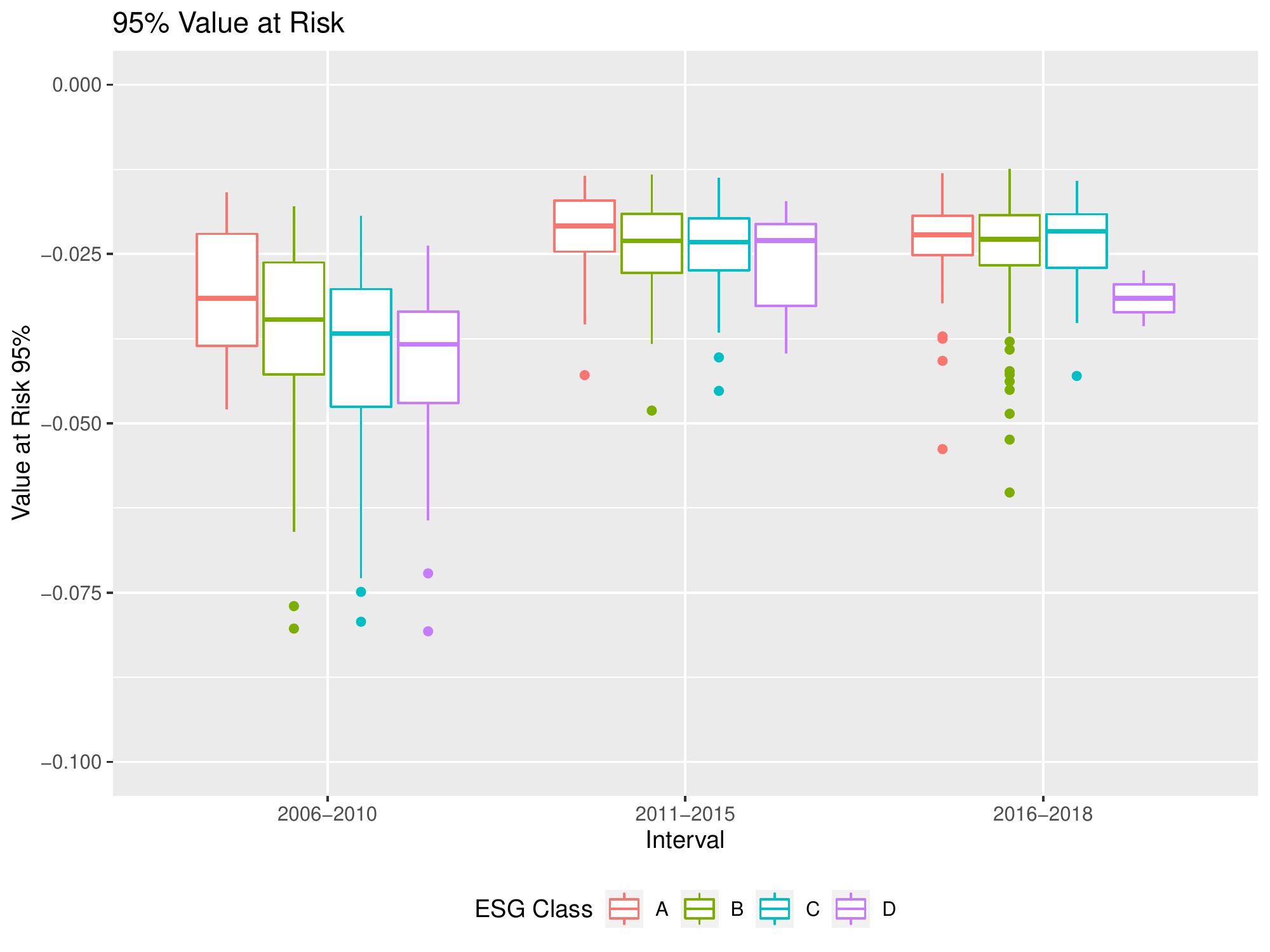}
\vspace{0.2cm}
\caption{Empirical 95\% Value at Risk classified by ESG class using Refinitiv thresholds and time period $q$.}
\label{VaR}
    \end{minipage}%
    \begin{minipage}{0.5\textwidth}
        \centering
\includegraphics[scale=0.38]{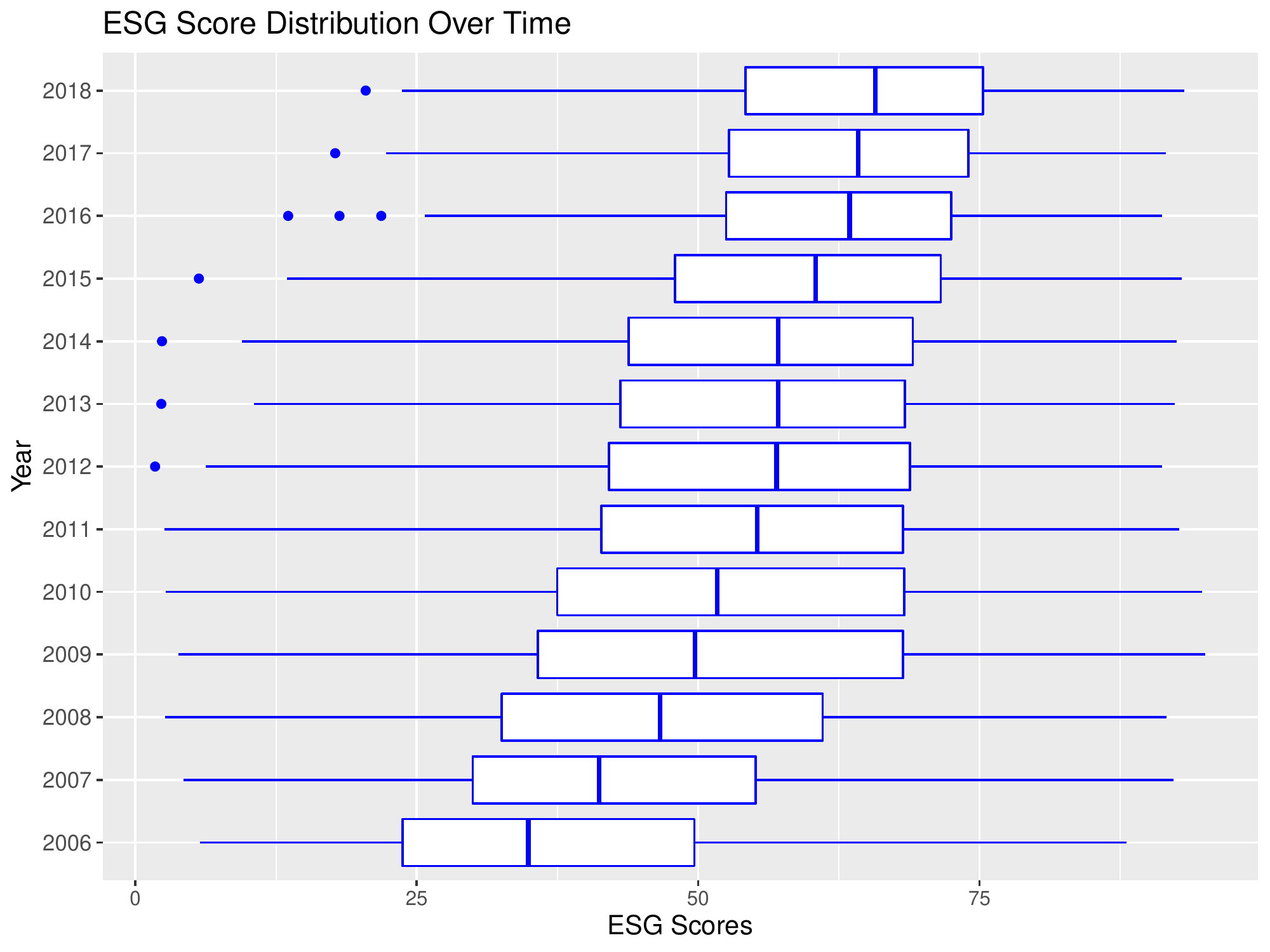}
\caption{Development of ESG scores over the years.}
\label{distribution}
    \end{minipage}
\end{figure}

To get a preliminary understanding if ESG scores can provide additional information on a company's riskiness,  the empirical 95\% Value at Risk (VaR), computed as the empirical quantile of the asset return distribution based on daily data for each asset $j$ in the period $q$ grouped according to their mean ESG scores across all sectors $S$ is presented in Figure \ref{VaR}.  The assets are attributed to the four different ESG classes \textit{A, B, C}, and \textit{D} following the thresholds (25, 50, 75) given by \cite{Refinitiv2021} as explained above.  Notice VaR values are smaller and exhibit larger variability during the time period 2006-2010, due to the 2008 financial crisis, while estimates become larger and, therefore,  less extreme in the other time periods. It is also quite evident, especially for the time periods 2006-2010 and 2011-2015, that ESG scores seem to be capable of providing information on the tail risk of an asset, as better-rated companies tend to have less negative median VaR values. This does not happen for 2016-2018, where  ESG scores are yet not definitive and might possibly be adapted in the future.  
Nevertheless, differences tend to diminish in the last two periods for classes \textit{A,}  \textit{B}, and \textit{C}  as also companies tend to improve their ESG scores, as shown in Figure \ref{distribution}. Still,  ESG class \textit{D}, which typically contain companies that have not yet fully disclosed ESG information and are also lower-rated than other companies, exhibit the worst VaR levels also for the time interval 2016-2018.   This is in line with the findings of \cite{Sahin2021} who show that there is no clear relationship with risk measures in the most recent ESG scores.

Summing up, empirical data suggest that ESG scores can provide information on tail riskiness of companies and allow to group assets that might share similar tail-risk characteristics, at least in the first two time periods. Plots for 99\% VaR and  95\% and 99\% Expected Shortfall show similar behaviour and are available upon request.

\section{R-Vine Copula ESG Risk Model and Methodology }

In general, a distributions rarely follows the strict spherical and elliptical assumptions as implied by correlation  \citep{ref:embrechts}. Therefore, according to \cite{ref:embrechts},  this traditional dependence measure is often not suited to create a proper understanding of the dependence in financial markets or fully comprehend the risk in extreme events.  While Pearson correlation has been used as a measure of pairwise dependence, modern risk management requires a thorough stochastic understanding beyond linear correlation.  As ESG risk is possible non-diversifiable (see \cite{Dorfleitner2016}), and the \cite{EBADiscussion} argues that ESG risk is supposed to manifest also in market risk, it takes a more complex risk measure to estimate its magnitude.  This is where copula modeling steps in.  



A copula $C$ is a cumulative distribution function (cdf) with univariate uniform marginal distributions on the unit interval. The copula approach is especially popular due to Sklar's theorem (\citeyear{ref:sklar}),  which allows to model the marginal distribution and the dependence separately.  
Therefore, if $F$ is a continuous $d$-dimensional distribution function of $\textbf{X} = (X_1 , \ldots , X_d)^\top$ with univariate cdf $F_p(x_p)$ of a continuous random variable $X_p$ for $p=1,\ldots, d$ with its realizations $x_p$,  the joint distribution function $F$ can be written as
\begin{equation}\label{lab:sklars}
F(x_1,\ldots,x_d) = C\bigg(F_1(x_1), \ldots, F_d(x_d)\bigg).
\end{equation}
The corresponding  density is
\begin{equation}\label{lab:sklars-density}
f(x_1,\ldots,x_d) = c\bigg(F_1(x_1), \ldots, F_d(x_d)\bigg)\cdot  \prod_{p=1}^{d} f_{p}(x_{p}),
\end{equation}
where $c$ is the $d$-dimensional copula density of the random vector $\textbf{F}=\big(F_1(X_1), \ldots, F_d(X_D)\big)^\top \in [0,1]^d$ and $f_p(x_p)$ is the associated univariate marginal density of $F_p(x_p)$ for $p= 1, \ldots, d$.

Different copula types with their reflections and rotations can accommodate flexible dependence patterns in the bivariate case ($d=2$), as shown in the Appendix in Table \ref{DifCop}.  Nevertheless, the existing parametric families of multivariate copulas are not as flexible as the bivariate copula families to represent complex dependence patterns. For instance, the multivariate Gaussian copula does not accommodate any tail dependence and has been strongly criticized after the 2008 financial crisis \citep{ref:1, ref:2, ref:puc, ref:czado1}. The multivariate Student's t copula allows for tail dependence but does not capture any asymmetry in the tails.  Furthermore,   multivariate exchangeable Archimedean copulas become inflexible as the dimension increases since they model the dependence between a large number of pairs of variables using not more than two parameters. If no dependence is found, and the random variables are independent, the independence copula best models their behavior.

In order to improve the copula method with regard to larger dimensions and to accommodate a great variety of dependence structures, vine copulas (so-called pair copula constructions) use conditioning and were made operational for data analysis by \cite{ref:aas}. They are a class of copulas, which are based on the conditioning ideas first proposed by \cite{ref:joe2} and further developed by \cite{ref:bedford2, ref:bedford1}. The approach allows to construct any $d$-dimensional copula and its density by $\frac{d\cdot(d-1)}{2}$ bivariate copulas and their densities. Furthermore, the expression can be represented by an undirected graphical structure involving a set of linked trees,  i.e., a \textit{regular (R-)vine structure} \citep{ref:bedford2}. An example is given in Figure \ref{fig:treeall} and more details on R-vines and vine trees are given by \cite{ref:kura, ref:kurojoe, ref:joe2014}; and \cite{ref:czado1}.

In previous works, copulas and vines copulas have had various financial applications.  Some examples include but are not limited to:  \cite{Bhatti2012} who applied time-varying copulas to capture the tail dependence between selected international stock and \cite{Nguyen2012} who used non-parametric and parametric copulas to capture the dependence between oil prices and stock markets.   Furthermore,  \cite{Naifar2012} modeled the dependence structure between risk premium, equity return,  and volatility in the presence of jump-risk, and \cite{ref:brechmann} analysed the sectorial dependence of Euro Stoxx 50.  Morover,  \cite{Fenech2015} discussed loan default correlation using an Archimedean copula approach, and \cite{Pourkhanali2016} used vine copulas to estimate systemic risk by looking at the connection of financial institutions. Recently,  \cite{Fink2017} and \cite{Ben2018} introduced the regime-switching vine copula approach,  and \cite{Abakah2021} re-examined international bond market dependence.  

\subsection*{R-Vine Copula ESG Risk Model}
\begin{figure}[H]

\begin{tikzpicture}[every node/.style = VineNode, node distance =1.4cm,scale=0.5, transform shape]

  
    \node (I_S) at (-10,10) {S\& P 500  $(M)$ };
    \node [left=of I_S ](I_A) {ESG Class Index A $(A)$};       
    \node [right=of I_S ] (I_C){ ESG Class Index C $(C)$ };
    \node [above=of I_S ] (I_B) {ESG Class Index B $(B)$ };    
      \node [below=of I_S ](I_D) {ESG Class Index D $(D)$ };

 \node [above left=of I_A ](A_n) {$a_{87}$};    
  \node [left=of I_A ](A_2) {$a_2$};    
 \node [below left=of I_A ](A_1) {$a_1$};    
      \node[TreeLabels] (T_1) [above of =A_2,rotate=50, yshift=-0.5cm]{$\cdots \cdots$};

  \node [above left=of I_B ](B_n) {$b_{85}$};    
  \node [above=of I_B ](B_2) {$b_2$};    
 \node [above right=of I_B ](B_1) {$b_1$};    
 \node[TreeLabels] (T_1) [above of =B_2] {$T_1$};
    \node[TreeLabels] (T_1) [left of =B_2,rotate=20]{$\cdots \cdots$};

 \node [above right=of I_C ](C_n) {$c_{84}$};    
  \node [right=of I_C ](C_2) {$c_2$};    
 \node [below right=of I_C ](C_1) {$c_1$};    
     \node[TreeLabels] (T_1) [above of =C_2,rotate=-45, yshift=-0.5cm]{$\cdots \cdots$};

  \node [below left=of I_D ](D_n) {$d_{78}$};    
  \node [below=of I_D ](D_2) {$d_2$};    
 \node [below right=of I_D ](D_1) {$d_1$};    
     \node[TreeLabels] (T_1) [left of =D_2,rotate=-20]{$\cdots \cdots$};

 %
	\draw (I_A) to node[draw=none, fill = none,  midway,  above,sloped, font=\scriptsize] {} (A_1);   
	\draw (I_A) to node[draw=none, fill = none,  midway,  sloped, above, font=\scriptsize] {} (A_2);   
	\draw (I_A) to node[draw=none, fill = none,  midway,  sloped, above, font=\scriptsize] {} (A_n);

	\draw (I_B) to node[draw=none, fill = none,  midway,  sloped, above, font=\scriptsize] {} (B_1);   
	\draw (I_B) to node[draw=none, fill = none,  midway,  sloped, above, font=\scriptsize] {} (B_2);   
	\draw (I_B) to node[draw=none, fill = none,  midway,  sloped, above, font=\scriptsize] {} (B_n);

	\draw (I_C) to node[draw=none, fill = none,  near end,  sloped, above, font=\scriptsize] {} (C_1);   
	\draw (I_C) to node[draw=none, fill = none,  midway,  sloped, above, font=\scriptsize] {} (C_2);   
	\draw (I_C) to node[draw=none, fill = none,  midway,  sloped, above, font=\scriptsize] {} (C_n);

	\draw (I_D) to node[draw=none, fill = none,  near end,  sloped, above, font=\scriptsize] {} (D_1);   
	\draw (I_D) to node[draw=none, fill = none,  midway,  sloped, above, font=\scriptsize] {} (D_2);   
	\draw (I_D) to node[draw=none, fill = none,  near end,  sloped, above, font=\scriptsize] {} (D_n);

	\draw (I_S) to node[draw=none, fill = none,  midway,  sloped, above, font=\scriptsize] {} (I_A);   
	\draw (I_S) to node[draw=none, fill = none,  midway,  sloped, above, font=\scriptsize] {} (I_B);   
	\draw (I_S) to node[draw=none, fill = none,  midway ,  sloped, above, font=\scriptsize] {} (I_C);   
     	\draw (I_S) to node[draw=none, fill = none,  midway ,  sloped, above, font=\scriptsize] {} (I_D);


\end{tikzpicture}
\hspace{4cm}
\centering
\parbox{7.5cm}{
\renewcommand{\xshiftNodes}{0.1*\linewidth}
\renewcommand{\yshiftLabels}{.0cm}  

\begin{tikzpicture}[every node/.style = VineNode, node distance =1.4cm,scale=0.5, transform shape]
    \node (I_B) at (-15,10) { $B$,$M$ };
    \node [left=of I_B ](I_A) {$A$,$M$ };       
    \node [right=of I_B ](I_C) {$C$,$M$ };       
    \node [right=of I_C](I_D) {$D$,$M$ };       

 \node [above left=of I_A ](A_n) {$a_{87}$, $A$};    
  \node [left=of I_A ](A_2) {$a_2$,$A$};    
 \node [below left=of I_A ](A_1) {$a_1$, $A$};    
     \node[TreeLabels] (T_1) [above of =A_2,rotate=50, xshift=-6, yshift=-6.5]{$\cdots$};

 \node [below left=of I_B ](B_n) {$b_{85}$, $B$};    
  \node [below=of I_B ](B_2) {$b_2$,$B$};    
 \node [below right=of I_B ](B_1) {$b_1$, $B$};    
      \node[TreeLabels] (T_1) [left of =B_2,rotate=-30, xshift=6, yshift=6.5]{$\cdots$};
 
 \node [above left=of I_C ](C_n) {$c_{84}$, $C$};    
  \node[above =of I_C ](C_2) {$c_2$,$C$};    
 \node [above right=of I_C ](C_1) {$c_1$, $C$};    
  \node[TreeLabels] (T_1) [above of =C_2, xshift=-2cm] {$T_2$};
     \node[TreeLabels] (T_1) [left of =C_2,rotate=30, xshift=6, yshift=-6.5]{$\cdots$};

  \node [above right=of I_D ](D_n) {$d_{78}$, $D$};    
  \node [right=of I_D ](D_2) {$d_2$, $D$};    
 \node [below right=of I_D ](D_1) {$d_1$, $D$};    
     \node[TreeLabels] (T_1) [above of =D_2,rotate=-50, xshift=8, yshift=-4.5]{$\cdots$};

 	\draw (I_A) to node[draw=none, fill = none,  midway, sloped, above,font=\scriptsize] {} (A_1);   
	\draw (I_A) to node[draw=none, fill = none,  midway,  sloped, above,font=\scriptsize] {} (A_2);   
	\draw (I_A) to node[draw=none, fill = none,  midway,  sloped, above,font=\scriptsize] {} (A_n); 
	  
 	\draw (I_B) to node[draw=none, fill = none,  midway, sloped, above,font=\scriptsize] {} (B_1);   
	\draw (I_B) to node[draw=none, fill = none,  midway,  sloped, above,font=\scriptsize] {} (B_2);   
	\draw (I_B) to node[draw=none, fill = none,  midway,  sloped, above,font=\scriptsize] {} (B_n);   

  	\draw (I_C) to node[draw=none, fill = none,  right, sloped, above,font=\scriptsize] {} (C_1);   
	\draw (I_C) to node[draw=none, fill = none,  midway,  sloped, above,font=\scriptsize] {} (C_2);   
	\draw (I_C) to node[draw=none, fill = none,  midway,  sloped, above,font=\scriptsize] {} (C_n);   
	
	 	\draw (I_D) to node[draw=none, fill = none,  midway, sloped, above,font=\scriptsize] {} (D_1);   
	\draw (I_D) to node[draw=none, fill = none,  midway,  sloped, above,font=\scriptsize] {} (D_2);   
	\draw (I_D) to node[draw=none, fill = none,  midway,  sloped, above,font=\scriptsize] {} (D_n);   

	\draw (I_A) to node[draw=none, fill = none,  midway,  sloped, above,font=\scriptsize] {} (I_B);   
	\draw (I_B) to node[draw=none, fill = none,  midway,  sloped, above,font=\scriptsize] {} (I_C);   
	\draw (I_C) to node[draw=none, fill = none,  midway ,  sloped, above,font=\scriptsize] {} (I_D);   

\end{tikzpicture}
}
\qquad
\begin{minipage}{7.5cm}
\renewcommand{\xshiftNodes}{0.1*\linewidth}
\renewcommand{\yshiftLabels}{.0cm}  
\vspace{1cm}
\begin{tikzpicture}[every node/.style = VineNode, node distance =1.4cm,scale=0.5, transform shape]

    \node (I_B) at (-15,10) { B,C ; M};
    \node [left=of I_B ](I_A) {A,B ;M};       
    \node [right=of I_B ](I_C) {C,D;M};       

 \node [above left=of I_A ](A_n) {$a_1$, M;A};    
 \node [below left=of I_A ](A_1) {$a_{87}$, M;A};    
    \node[TreeLabels] (T_1) [below of= A_1, rotate=90, xshift=3cm] {$\cdots \cdots $};

  \node (B_n)at (-13, 12) {$b_1$, M;B};    
 \node (B_1)  at (-16.5, 12){$b_{85}$, M;B};    
      \node[TreeLabels] (T_1) [right of= B_1, xshift=0.4cm] {$\cdots \cdots $};

   \node[TreeLabels] (T_1) [above of =B_1, xshift=1.5cm] {$T_3$};

   \node (C_n) at (-13, 8){$c_1$, M;C};    
 \node (C_1) at (-16.5,8) {$c_{84}$, M;C};    
  \node [below right =of I_C](D_1) {$d_{78}$, M;D};    
     \node[TreeLabels] (T_1) [right of= C_1, xshift=0.4cm] {$\cdots \cdots $};
     
    \node [above right =of I_C](D_n) {$d_1$, M;D};    
 \node [below right =of I_C](D_1) {$d_{78}$, M;D};    
     \node[TreeLabels] (T_1) [below of= D_1, rotate=90, xshift=3cm] {$\cdots \cdots $};

 	\draw (I_A) to node[draw=none, fill = none,  midway, sloped, above,font=\scriptsize] {} (A_1);   

	\draw (I_A) to node[draw=none, fill = none,  midway,  sloped, above,font=\scriptsize] {} (A_n); 
	  
 	\draw (I_B) to node[draw=none, fill = none,  midway, sloped, above,font=\scriptsize] {}  (B_1);   

	\draw (I_B) to node[draw=none, fill = none,  midway,  sloped, above,font=\scriptsize] {}  (B_n);

  	\draw (I_B) to node[draw=none, fill = none,  right, sloped, above,font=\scriptsize] {}  (C_1);   

	\draw (I_B) to node[draw=none, fill = none,  midway,  sloped, above,font=\scriptsize] {}  (C_n);   

  	\draw (I_C) to node[draw=none, fill = none,  right, sloped, above,font=\scriptsize] {}  (D_1);   

	\draw (I_C) to node[draw=none, fill = none,  midway,  sloped, above,font=\scriptsize] {}  (D_n);

	\draw (I_A) to node[draw=none, fill = none,  midway,  sloped, above,font=\scriptsize] {} (I_B);   
	\draw (I_B) to node[draw=none, fill = none,  midway,  sloped, above,font=\scriptsize] {}  (I_C);   

\end{tikzpicture}
\vspace{1cm}
\end{minipage}

\centering
\parbox{7.5cm}{
\renewcommand{\xshiftNodes}{0.1*\linewidth}
\renewcommand{\yshiftLabels}{.0cm}

\begin{tikzpicture}[every node/.style = VineNode, node distance =1.4cm,scale=0.6, transform shape]

    \node (I_B) at (-15,10) { D,B ;M,C};
    \node [left=of I_B ](I_A) {A,C ; M, B};       

 \node [below left=of I_A ](A_n) {$a_1$,B;A,M};    
 \node [below =of I_A ](A_1) {$a_{87}$,B;A,M};     
    \node[TreeLabels] (T_1) [above left of= A_n, xshift=3.2cm, rotate=-35,yshift=-2cm] {$\cdots$};

  \node [above left =of I_A)](B_n) {$b_1$,C;B,M};    
 \node [above  =of I_A ](B_1) {$b_{85}$,C;B,M};    
    \node[TreeLabels] (T_1) [above left of= B_n, xshift=2cm, rotate=15,yshift=-0.5cm] {$\cdots$};

  \node[TreeLabels] (T_1) [above of= B_n, xshift=4.5cm] {$T_4$};

 \node [above right=of I_B ](C_n) {$c_1$,B;C,M};    
 \node [above =of I_B ](C_1) {$c_{84}$B;C,M};     
    \node[TreeLabels] (T_1) [above right of= C_1, xshift=1.2cm, rotate=-35,yshift=-1.5cm] {$\cdots$};

  \node [below right =of I_B)](D_n) {$d_1$,C;D,M};    
 \node [below  =of I_B ](D_1) {$d_{78}$,C;D,M};    
    \node[TreeLabels] (T_1) [above right of= D_1, xshift=0.3cm, rotate=20,yshift=-1.0cm] {$\cdots$};

 	\draw (I_B) to node[draw=none, fill = none,  midway, sloped, above,font=\scriptsize] {} (C_1);   

	\draw (I_B) to node[draw=none, fill = none,  midway,  sloped, above,font=\scriptsize] {} (C_n); 
	  
 	\draw (I_B) to node[draw=none, fill = none,  midway, sloped, above,font=\scriptsize] {}  (D_1);   

	\draw (I_B) to node[draw=none, fill = none,  midway,  sloped, above,font=\scriptsize] {}  (D_n);

 	\draw (I_A) to node[draw=none, fill = none,  midway, sloped, above,font=\scriptsize] {} (A_1);   

	\draw (I_A) to node[draw=none, fill = none,  midway,  sloped, above,font=\scriptsize] {} (A_n); 
	  
 	\draw (I_A) to node[draw=none, fill = none,  midway, sloped, above,font=\scriptsize] {}  (B_1);   

	\draw (I_A) to node[draw=none, fill = none,  midway,  sloped, above,font=\scriptsize] {}  (B_n);

	\draw (I_A) to node[draw=none, fill = none,  midway,  sloped, above,font=\scriptsize] {} (I_B);

\end{tikzpicture}
}
\qquad
\begin{minipage}{7.5cm}
\renewcommand{\xshiftNodes}{0.1*\linewidth}
\renewcommand{\yshiftLabels}{.0cm}

\begin{tikzpicture}[every node/.style = VineNode, node distance =1.7cm,scale=0.5, transform shape]

    \node (I_B) at (-18,10) { A,D ; M, B,C};

 \node (A_n)  at (-23, 12)  {$a_1$,C;A,M,B};    
 \node (A_1)at (-23, 9) {$a_{87}$,C;A,M,B};    
         \node[TreeLabels] (T_1) [below of= A_1, rotate=90, xshift=3cm] {$\cdots \cdots $};
         
  \node (B_n)at (-20,13) {$b_1$,A;B,M,C};    
 \node (B_1) at (-20,7) {$d_1$,B;D,M,C};    
   \node[TreeLabels] (T_1) [above of= B_n, xshift=1.5cm] {$T_5$};
          \node[TreeLabels] (T_1) [right of= B_1] {$\cdots $};
          \node[TreeLabels] (T_1) [right of= B_n] {$\cdots $};

 \node (C_n) at (-13. 5,12) {$c_1$, D;C,M,B};    
 \node (C_1) at (-13.5, 9) {$c_{84}$D;C,M,B};    
          \node[TreeLabels] (T_1) [below of= C_1, rotate=90, xshift=3cm] {$\cdots \cdots $};
          
  \node (D_n)at  (-16.5,13) {$b_{85}$,A;B,M,C};    
 \node (D_1) at (-16.5, 7) {$d_{78}$,B;D,M,C};

 	\draw (I_B) to node[draw=none, fill = none,  midway, sloped, above,font=\scriptsize] {} (C_1);   

	\draw (I_B) to node[draw=none, fill = none,  midway,  sloped, above,font=\scriptsize] {} (C_n); 
	  
 	\draw (I_B) to node[draw=none, fill = none,  midway, sloped, above,font=\scriptsize] {}  (D_1);   

	\draw (I_B) to node[draw=none, fill = none,  midway,  sloped, above,font=\scriptsize] {}  (D_n);

 	\draw (I_B) to node[draw=none, fill = none,  midway, sloped, above,font=\scriptsize] {} (A_1);   

	\draw (I_B) to node[draw=none, fill = none,  midway,  sloped, above,font=\scriptsize] {} (A_n); 
	  
 	\draw (I_B) to node[draw=none, fill = none,  midway, sloped, above,font=\scriptsize] {}  (B_1);   

	\draw (I_B) to node[draw=none, fill = none,  midway,  sloped, above,font=\scriptsize] {}  (B_n);

\end{tikzpicture}

\end{minipage}
\caption{The first five vine trees of the regular vine used.}
\label{fig:treeall}
\end{figure}
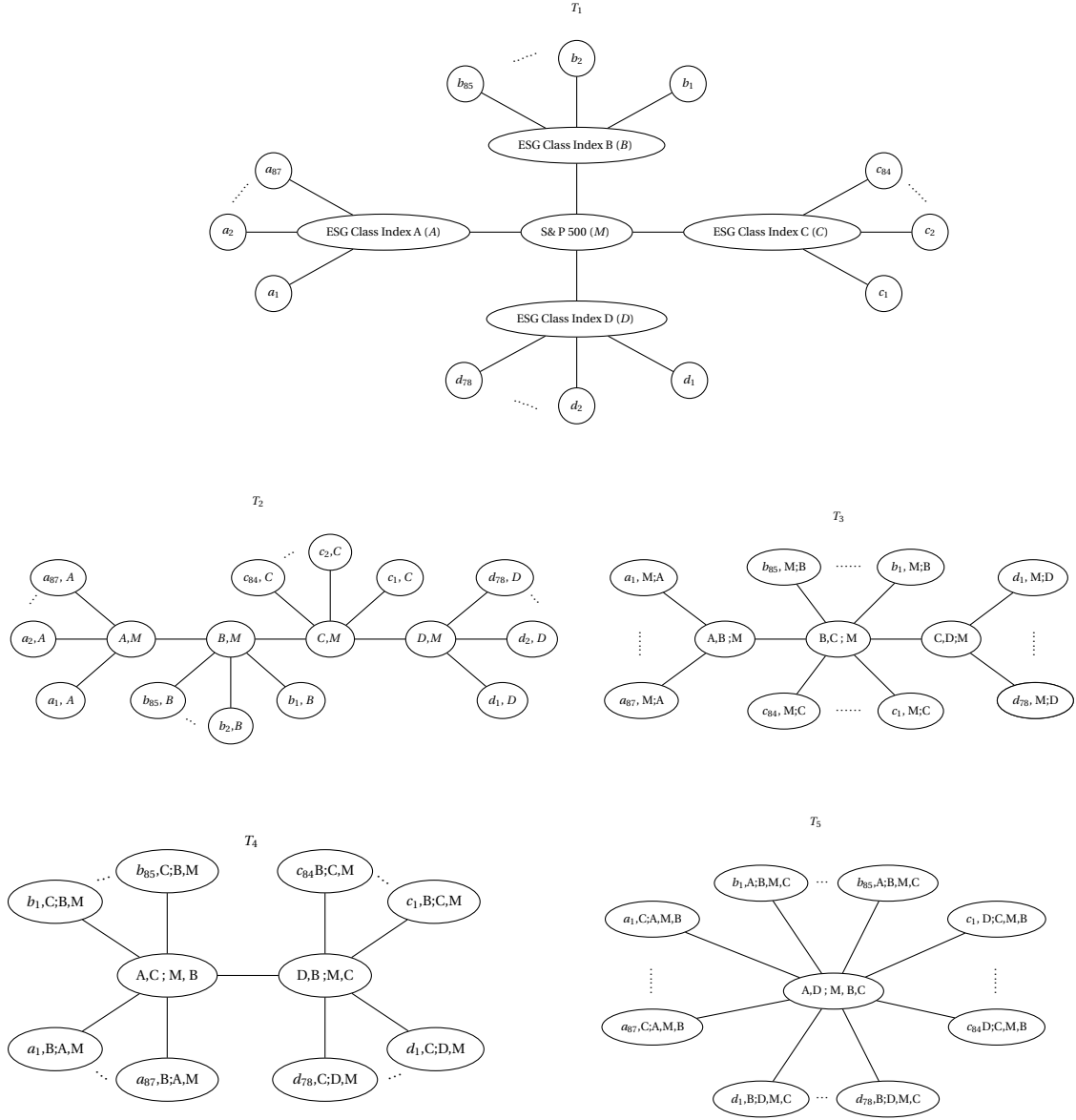

In this research,  we apply a R-vine copula model to  estimate different dependence structures among assets given their ESG classes in order to compute different risk measures. To  be able to capture the ESG risk and market risk conditionally on the ESG class of an asset $j$, the proposed first five vine tree structures are defined as shown in Figure \ref{fig:treeall}. 
 In the first vine tree $T_1$ in Figure \ref{fig:treeall}, we connect the assets $j$ (abbreviated by \sloppy $a_1, \cdots, a_{87}, b_1, \cdots, b_{85}, c_1, \cdots, c_{84}, d_1, \cdots, d_{78}$) to their belonging ESG class index ($I_{t,k}^{q} $ abbreviated by $k \in \{A,B,C,D\}$) and market index ($M_t^{q}$ abbreviated by $M$) in period $q$. This allows us to compute the dependence of  an asset $j$ with its ESG class, later we use this to compute its ESG risk in Equation (\ref{riskintext1}). In $T_2$, in Figure \ref{fig:treeall}, we connect the nodes so that they allow us to compute the market risk conditionally on the ESG class (see  Equation (\ref{riskintext2})) following the sequential ESG class order.  We then continue to fix  $T_3$ to $T_5$ to be able to compute our risk measures as a  fraction of all dependence, positive and negative, an asset has with all ESG classes $k$.
In $T_3$  in Figure \ref{fig:treeall} we choose to connect the nodes so that we separate the best and lowest ESG performance from the middle ESG performance, classes $B$ and $C$.  By defining $T_4$ and $T_5$  further, we guarantee that in our risk measures in Section \ref{riskmeasures},  only dependencies with the different ESG class indices $I_{t,k}^q$ are included, which is not yet modelled in $T_1$, $T_2$, and $T_3$. This ensures that the ESG risk is not an individual measure and takes into account complex market movements and dependencies with other ESG classes $k$.

\subsection*{Inputs for the R-Vine Copula ESG Risk Model }
In order to be able to fit the R-vine copula ESG risk model with the five specified vine trees,  the inputs have to be computed.  We start by computing the mean ESG score  ($\overline{ESG}_j^{q}$) for each asset $j$ for each time period  $q$.  We then divide the assets into their ten different economic sectors and rank them according to their $\overline{ESG}_j^{q}$.  This allows us to group the assets within each sector  into four different quartiles according to their ESG performance.  Within each sector, we categorize assets with  the highest scores in the first quartile as ESG class $A$,  assets with the second to highest scores in the second quartile as ESG class $B$,  assets with the second to lowest scores in the third quartile, as ESG class $C$, and finally, assets with the lowest scores in the fourth quartile, as ESG class $D$ (for details see  \ref{mean} and \ref{define}).

We define the ESG class quartiles as $k \in (A,B,C,D)$. We choose to rank the assets within their economic sector instead of overall as it has been shown by \cite{Kumar2016} that ESG scores affect various sectors to a different degree.  Furthermore, we know that ESG scores values are weighted differently within economic sectors \citep{Refinitiv2021}.  Therefore, categorizing the assets within their sector allows us to take into account these differences.  More importantly, it  also allows us to later compute the ESG risk measures that are comparable across sectors.  As it has been shown that that  investors cannot eliminate ESG risk fully through diversification (see \cite{Dorfleitner2016}), it is not sufficient to compute the risk on an individual level without taking a look at the surrounding assets.  We choose to use quartiles instead of Reuters's thresholds to avoid low numerosity in some classes. 
We now combine all assets according to their  ESG class quartile  $k$. This means that each ESG rating class $k$ includes assets from all ten different sectors.   Overall, ESG class $A$ includes 87 assets, ESG class $B$ includes 85 assets,  ESG class $C$ includes 84 assets, and finally ESG Class $D$ includes $78$ assets.   

Next, we compute the ESG class indices per ESG class $k$ ($I_{t,k}^{q}$, i.e.  the ESG class index \textit{A} in period $q$ on trading day $t$ is $I_{t,A}^{q}$) by linearly combining asset returns ($Y_{t,j,k}^{q}$ defined as the return of an asset $j$  in period $q$ on trading day $t$ belonging to ESG class $k$) in the same mean ESG class ($K_{j}^{S,q}$) using the market capitalization weights.  As shown in Figure \ref{fig:treeall} we use only the variable names such as $M$ (Market), $A$ (ESG class index), $a_1$ (asset) etc.  to denote the nodes. The associated daily data is given by $I_{M,t}^q$, $I_{t,A}^{q}$, $Y_{t,1,A}^{q}$ and a full table of notation can be found in \ref{Information_Notation}.
  
Then, we use the two step inference for margins approach,  proposed by \cite{JoeIFM}, to compute our pseudo-copula data which serves as the input for our R-vine copula ESG risk model (for details see \ref{IFM}) and allows us to estimate the copula parameters of the chosen bivariate copula family (for details see \ref{Familyselection}).



Additionally, when fitting the R-vine copula ESG risk model, we can specify different bivariate copula families for the pairs of variables.  In the first specification, we allow only for the bivariate \emph{itau} copulas, for which the estimation by Kendall’s $\tau$ inversion is available (Student's t, Frank, Gaussian, Clayton,  Joe,  Gumbel,  and Independence copula). Using these,  we also allow for asymmetry in the upper and lower tails, as shown in Table \ref{DifCop} in the Appendix. These copula families have a one-to-one relationship between their copula parameter and Kendall’s $\tau$. In the second specification, we include additional copula families with two parameters such as BB1 (a combination between both extreme cases of Clayton copula and Gumbel copula), BB7 (a combination of Joe copula and Clayton copula), and the extreme-value copula BB8 (extended Joe) in order to see if this improves our model fit.  
 Lastly, we fit a specification using only Gaussian copulas to see the change in model fit if we do not account for the dependence in the extremes. To choose our optimal bivariate copula family specification, we use the modified Bayesian Information Criteria (mBIC) instead of the widely used Bayesian Information Criteria (BIC) by \citep{Schwarz1978}.  The mBIC has been proposed by \cite{Nagler2019} and is tailored to sparse vine copula models in high-dimensions.  These scholars show that the mBIC can consistently distinguish between the true and alternative models.  We then continue our analysis using the best fitting model only.


\subsection*{R-Vine Risk Measures}\label{riskmeasures}

After computing the inputs for the R-vine copula ESG risk model,  estimating the pseudo-copula data, and fitting it to our model with the specified five vine trees, we can compute three different risk measures using the overall dependence $(\tau)$ and lower (tail) dependence $(\lambda)$. 
We  introduce an overall ESG risk $R_{j_k^q}^{ESG}(\tau)$ and lower tail ESG risk $R_{j_k^q}^{ESG}(\lambda)$, which we can compute for each asset $j$ with its ESG class $k$ in period $q$. It accounts for the ESG risk of an asset $j$ belonging to a specific ESG class $k$ and gives the fraction of all dependence, positive and negative, explained by the assets ESG class. Additionally, we estimate the market risk conditionally on the ESG class for each asset $j$ as $R_{j_k^q}^{Market}(\tau)$ and  $R_{j_k^q}^{Market}(\lambda)$ with all the (tail) dependence on the ESG class being removed. Lastly, by subtracting both indicators from 1, we can get the overall idiosyncratic risk $R_{j_k^q}^{idio}(\tau)$ and lower tail idiosyncratic risk as $R_{j_k^q}^{idio}(\lambda)$ for each asset $j$ within each time period $q$.  See \cite{ref:brechmann} for sectorial estimation.

We choose the empirical Kendall’s $\tau$ and its estimate  $\hat{\tau}$, whose values range from $[-1, 1]$ \citep{ref:joe2014}, as our dependence measure for the overall risk measures\footnote{
	Kendall's $\tau$ is a ranked based dependence measure robust to outliers that fits the aim of our analysis and can be defined  in terms of copulas for two continuous random variables $(X_1,X_2)$ with copula $C$ as $\tau = 4  \int \limits_{[0,1]^2} C(u_1, u_2) dC(u_1, u_2)-1$. An alternative overall weight measure could be Spearman’s $\rho$.   }  :
	$R_{j_k^q}^{ESG}(\tau)$, $R_{j_k^q}^{Market}(\tau)$, and $R_{j_k^q}^{idio}(\tau)$. To exclude unrealistic assumptions, we assume that asset \textit{j} is not independent of all other assets and include all Kendall's $\tau$ only in absolute value.  Similarly, we can also define these risk measures using the lower tail dependence coefficient\footnote{When Student's t and Clayton (and BB1 and BB7) copulas are chosen as best fit among the bivariate copula families given in Table \ref{DifCop}, we can use this information to also estimate the lower tail dependence coefficient $\lambda$ which ranges from $[0, 1]$ and is defined for a bivariate distribution with copula $C$  as  $ \lambda^{lower} = \lim_{x \to 0^+}  P(X_2 \leq F_2^{-1}(t)\mid X_1 \leq F_1^{-1}(t)) = \lim_{x \to 0^+}  \dfrac{C(t,t)}{t}$. The lower tail dependence coefficient is also non-zero when the fitted bivariate copula class is Student's \textit{t}, Clayton, 180° Joe,  180° Gumbel, BB1,  BB7, or 180° BB8 in our model.  For other bivariate copula families, we have zero lower tail dependence coefficient.  Other methods could include estimating a specific distribution or a family of distributions; or working with a non-parametric model.  We refer to \cite{ref:frahm} for further reading on estimation methods.}
	 as : $R_{j_k^q}^{ESG}(\lambda)$, $R_{j_k^q}^{Market}(\lambda)$, and $R_{j_k^q}^{idio}(\lambda)$. By using the lower tail dependence coefficient as the input value,  these risk measures quantify the strength of the dependence within the lower-left-quadrant tail of an assets return and its associated ESG class index in relation to all other assets within the fitted R-vine copula ESG risk model.  The values for the risk measures are in the interval [0, 1].  As the equations are identical apart from dependence measure $\tau$ or $\lambda$,  only the overall risk measures are presented below.  
For simplification, we drop the sector index $S$ in the notation of the risk measures for each asset $j$.

\subsubsection*{Overall ESG Risk }

For each asset of ESG Class $A$ in period $q$, i.e.  $j_A^{q} \in \{ j |  K_{j}^{S,q} =A$ for $q=1,2,3 $ and $S=1, \cdots, 10 \}$ and $q=1,2,3$:
\begin{equation}\label{riskintext1}
R_{j_A^q}^{ESG}(\tau) = \frac{|\hat{\tau}_{j_A^q,I_A^q}|}{|\hat{\tau}_{j_A^q,I_A^q}|+|\hat{\tau}_{j_A^q,I_M^q|I_A^q}|+|\hat{\tau}_{j_A^q,I_B^q|I_A^q,I_M^q}|+|\hat{\tau}_{j_A^q,I_C^q|I_A^q,I_M^q,I_B^q}|+|\hat{\tau}_{j_A^q,I_D^q|I_A^q,I_M^q,I_B^q,I_C^q}|
}.
\end{equation} 

$R_{j_B^q}^{ESG}(\tau)$, $R_{j_C^q}^{ESG}(\tau)$, and $R_{j_D^q}^{ESG}(\tau)$ can be derived similarly.

\subsubsection*{Overall Market Risk conditionally on ESG Class}

For each asset of ESG Class $A$ in period $q$, i.e. $j_A^{q} \in \{ j |  K_{j}^{S,q} =A$ for $q=1,2,3  $ and $S=1, \cdots, 10 \}$ and  $q=1,2,3$:
\begin{equation}\label{riskintext2}
R_{j_A^q}^{Market}(\tau) = \frac{|\hat{\tau}_{j_A^q,I_M^q|I_A^q}|}{|\hat{\tau}_{j_A^q,I_A^q}|+|\hat{\tau}_{j_A^q,I_M^q|I_A^q}|+|\hat{\tau}_{j_A^q,I_B^q|I_A^q,I_M^q}|+|\hat{\tau}_{j_A^q,I_C^q|I_A^q,I_M^q,I_B^q}|+|\hat{\tau}_{j_A^q,I_D^q|I_A^q,I_M^q,I_B^q,I_C^q}|
}.
\end{equation} 

$R_{j_B^q}^{Market}(\tau)$, $R_{j_C^q}^{Market}(\tau)$, and $R_{j_D^q}^{Market}(\tau)$ can be derived similarly.

\subsubsection*{Overall Idiosyncratic Risk}
For each asset of ESG Class $A$ in period $q$, i.e. $j_A^{q} \in \{ j |  K_{j}^{S,q} =A$ for $q=1,2,3  $ and $S=1, \cdots, 10 \}$ and  $q=1,2,3$:

\begin{equation}\label{lab:Class}
R_{j_A^q}^{idio}(\tau)= 1-R_{j_A^q}^{ESG}(\tau)-R_{j_A^q}^{Market}(\tau).
\end{equation} 
$R_{j_B^q}^{idio}(\tau)$,$R_{j_C^q}^{idio}(\tau)$ and $R_{j_D^q}^{idio}(\tau)$  can be derived similarly.

Summing up, fitting the R-vine copula ESG risk model enables us to capture complex dependence structures as it allows us to account for asymmetric and tail dependence.  Moreover, to get a more comprehensive understanding of the overall share of dependence an asset has with other assets within the same ESG class given all other assets, these (tail) risk measures are standardized as ratios  instead of single dependence measures of an asset \textit{j} with its ESG class index alone. This helps to understand if there are common behaviors - especially for tail dependence.  Common dependencies could indicate that a time series (e.g., returns) exhibit co-movements and, therefore, could share some risk properties. As many investors tilt their portfolio towards comparable large ESG score, using an inclusion or exclusion approach,  understanding the dependence between similarly ESG rated assets is necessary in order to be able to capture co-movements and promote diversification.

\section{Empirical Results}\label{sec:Results}
After fitting the R-vine copula ESG risk model for three copula family specifications (\emph{itau, parametric, Gaussian}), we find that according to the mBIC, the  R-vine copula ESG risk model with the \emph{itau} copula family specification best fits our model and allows to most optimally estimate the dependencies between the assets for all three time periods $q$. We also notice that the\textit{ Gaussian} R-vine copula ESG risk model always performs worst.  Also, when looking at which of the copula families are fitted most often within each specification, we find that not very often the Gaussian copula is chosen for the \emph{itau} and \emph{parametric} R-vine copula ESG risk model. 
This is especially true for the first vine tree $T_1$ which models the dependence of an asset with its ESG class.  As the Gaussian copula does not capture any tail dependence, and often the Student's t copula is chosen, these results indicate that the assets definitely share some tail dependence. Thus,  the \textit{Gaussian} R-vine copula ESG risk model does not fully comprehend the risk in extreme events.   In our analysis, we therefore use the \emph{itau} R-vine copula ESG risk model to estimate our risk measures.  For more details, the information criteria for all three models in each time period $q$ are given in the \ref{Modelfit} and the specific number of copula families fitted in the first vine tree $T_1$ are given in \ref{Families}, while all other vine trees are available upon author request.

\begin{figure}[H]
\centering
\includegraphics[scale=0.5]{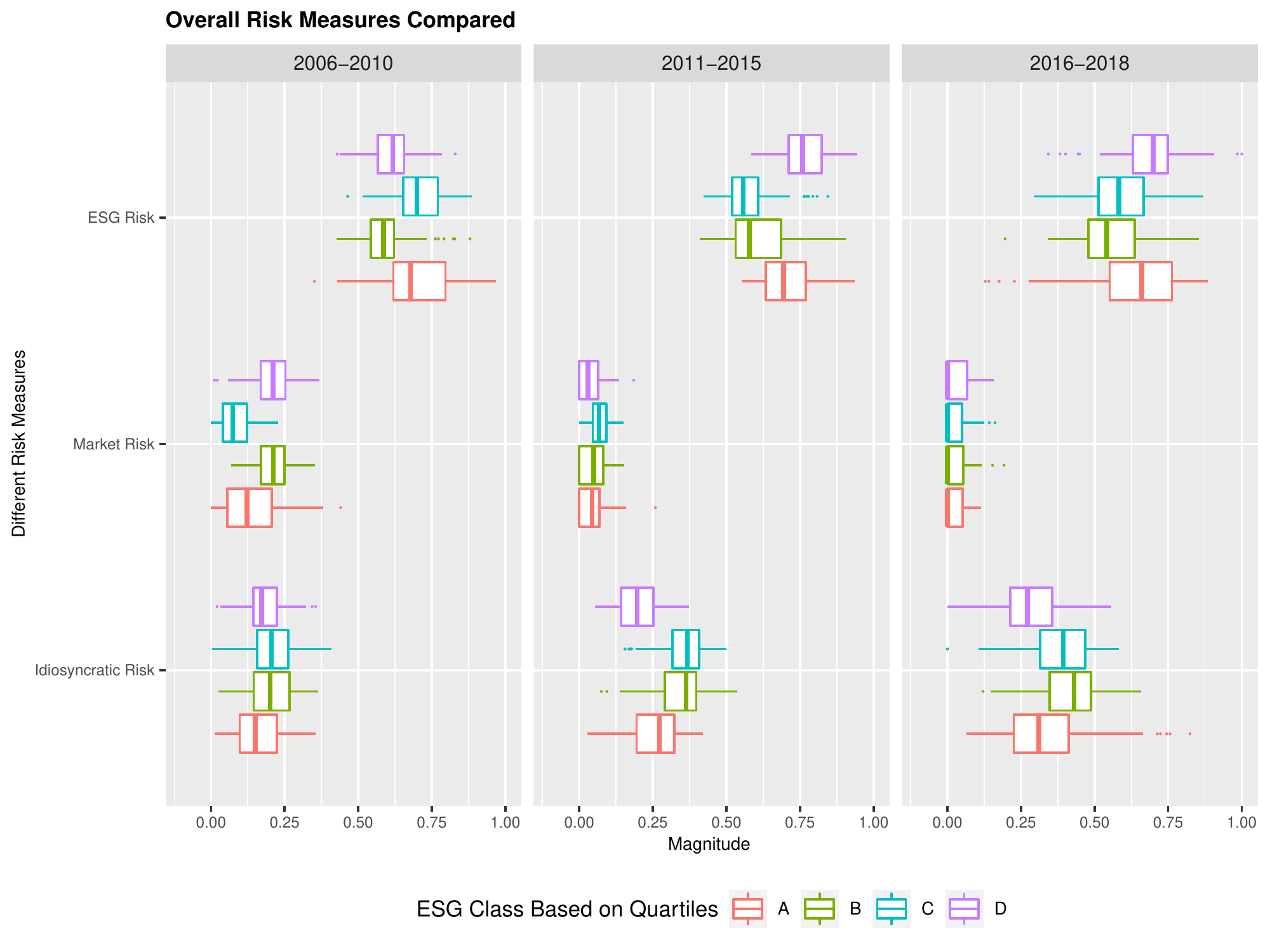}
\caption{Overall ESG risk measures for each time interval $q$ using quartiles for ESG classes. }
\label{Overall}
\end{figure}

In Figure \ref{Overall} and Table \ref{TableOverall} focusing on overall ESG risk we find that assets that belong to ESG class $D$ perform the worst, as they experience the highest overall ESG risk in calm periods (2011-2015, 2016-2018).  This is in line with the literature, as \cite{ref:shafer} argue that strong ESG practices can act as insurance against left-tail events as well as others who find that superior ESG performance reduces volatility \citep{Albuquerque2020, Bouslah2018}.  In contrast to other scholars who found that ESG performance mitigates financial risk during crisis using Chinese companies and the Covid-19 crisis \cite{Broadstock2021}, we find that in times of the financial crisis 2006-2010,  especially assets belonging to a $B$ ESG class tend to have the lowest overall ESG risk.  Assets belonging to  ESG class $A$ show unexpected worse overall ESG risk, possibly because of a high investment volume and popularity from investors who wrongly believe that ESG performance can make them resilient in times of crisis.  These findings are in line with \cite{Demers2021} who argue that ESG scores did not immunize stocks during the COVID-19 crisis and did, therefore, not protect the investors from unexpected losses.  Furthermore,  \cite{Flori2019}, who look at bipartite network representation of the relationships between mutual funds and portfolio holdings, find that the popularity of assets does not necessarily yield a beneficial outcome.  This shows that portfolios which invest in less popular assets generally outperform those investing in more popular ones.  Especially in the most recent time period, we find a larger variability of overall ESG risk belonging to the different classes.  All standard deviation values can be found in \ref{TableOverall_Std}. Companies that are not best or worst performers, i.e.  assets in ESG classes $B$ and $C$, show the lowest overall ESG risk.  Again, this could be due to the non-popularity of assets with mediocre ESG performance, as many investment strategies focus on inclusion or exclusion approaches. We also have to keep in mind that the ESG scores are not yet definitive; therefore, they could still be changed and adapted to change the ESG risk behaviour.  

When looking at the overall market risk conditionally on the ESG class, we find it to be relatively low for all assets. The difference among ESG classes also diminishes in times of calm and in the most recent period. Nevertheless, in times of crisis, 2006-2010, assets in the ESG class $A$ seem to carry the least market risk, which is in line with the literature as ESG practices have been connected to better governance and possible reduction in risk exposure. 

 Finally, the overall idiosyncratic component of the assets varies again throughout the time periods. In times of crisis, companies with good ESG performance show lower overall idiosyncratic risks,  while companies with a lower ESG score show worse risks.  This is in line with the recent literature, as \cite{Becchetti2015} show that ESG investing reduces the idiosyncratic volatility exposure. This behaviour changes when looking at times of calm in 2011-2015 and the most recent period. Here companies that either have very high or very low ESG scores indicate the lowest idiosyncratic risk.     However, this does not necessarily mean that including the extremes in a portfolio results in the best possible performance as  \cite{Campbell2001} and \cite{Luo2009} have shown that the realized performance in portfolios depends on the overlapping effects of systematic and idiosyncratic risks.

\begin{table}[H]
\centering
\footnotesize
\begin{tabular}{c|cccc|cccc|cccc}
\hline
Type of Risk &  \multicolumn{4}{c|}{ESG Risk}     & \multicolumn{4}{c|}{Market Risk}   & \multicolumn{4}{c}{Idiosyncratic Risk}    \\
\hline
Year         & A        & B    & C    & D    & A           & B    & C    & D    & A                  & B    & C    & D   \\
\hline
2006-2010    & 0.696          & 0.589 & 0.706 & 0.614 & 0.138             & 0.205 & 0.083 & 0.207 & 0.166                    & 0.206 & 0.211 & 0.179 \\
2011-2015   & 0.704          & 0.611 & 0.577 & 0.766 & 0.044             & 0.052 & 0.066 & 0.039 & 0.252                    & 0.338 & 0.357 & 0.195 \\
2016-2018    & 0.630          & 0.564 & 0.595 & 0.693 & 0.027             & 0.029 & 0.027 & 0.030 & 0.343                    & 0.407 & 0.378 & 0.276  \\
 \hline
\end{tabular}
\caption{Mean values for overall ESG Risk, Market Risk and Idiosyncratic Risk for each ESG Class $k$ in time interval $q$.}
\label{TableOverall}
\end{table}

\begin{figure}[H]
\centering
\includegraphics[scale=0.5]{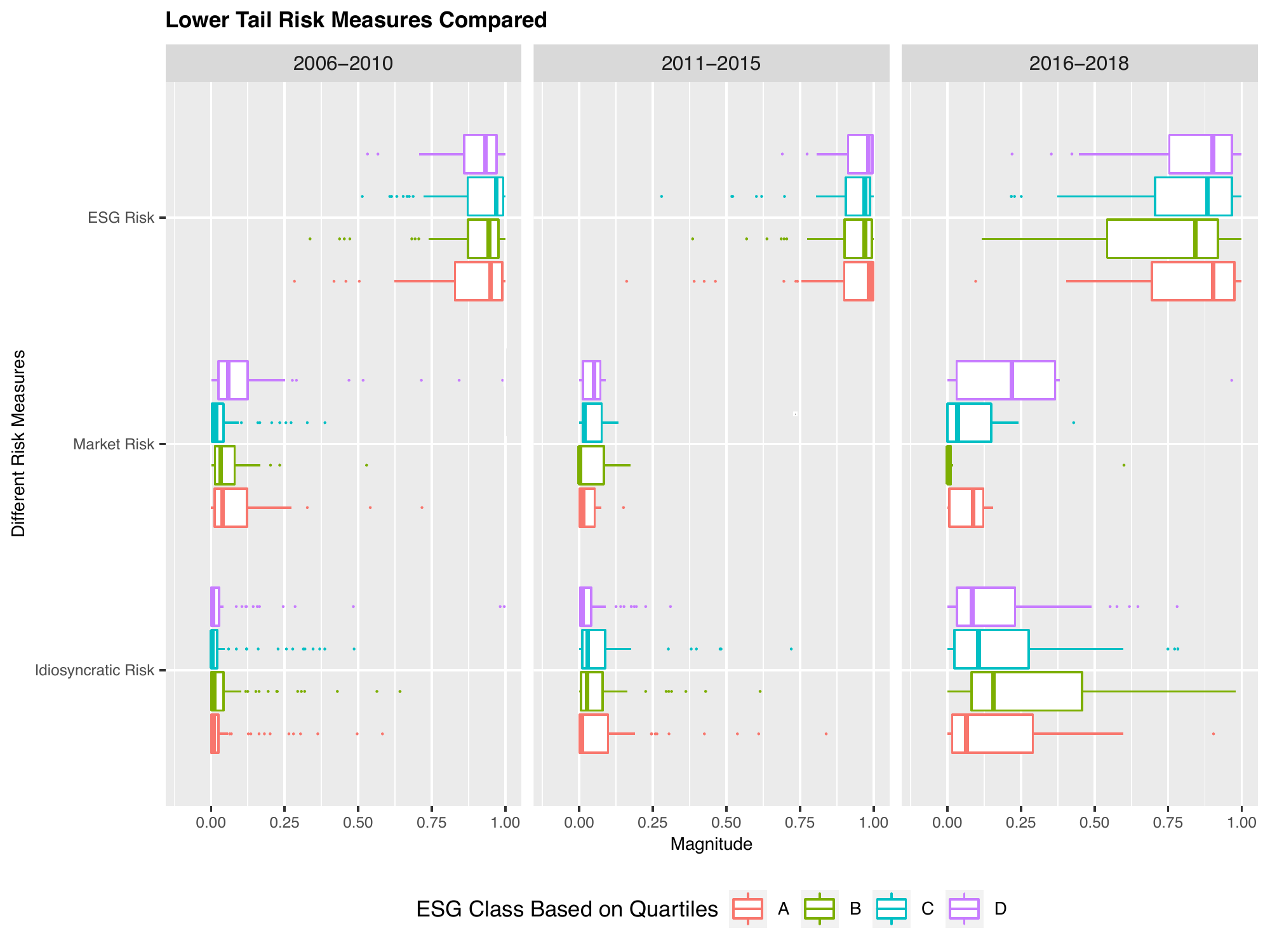}
\caption{Lower tail ESG risk measures (without 0s and 1s) for each time interval $q$ using quartiles for ESG classes.}
\label{lower}
\end{figure}

When looking at the lower tail risk measures in Figure \ref{lower} and Table \ref{tablelower}, using the estimated lower tail dependence coefficient ($\lambda$), we first notice that the magnitude is close to 1. This is due to the design of the indicator and the number of copula families with zero tail dependence coefficients. Nevertheless, we can still compare the risk measures across ESG classes and time periods.  The intuition why companies with high ESG scores could potentially encounter lower downside risks is that it is common believe that socially responsible companies are less exposed to company-specific events that negatively impact the equity price \citep{Dietmont2016}.  Using questionnaires and annual assessments,  some scholars already found a significant relationship between certain aspects of CSR and downside tail risk which differs when looking at the region and time \citep{Dietmont2016}.

We find that assets with ESG class $A$ show the most favourable lower tail ESG risk throughout the first two time periods.  However, the difference diminishes in 2011-2015 and changes in the most recent period. While some scholars argue that ESG tail risk has become more pronounced in the most recent year using the Covid-19 crisis \citep{new:loof2021},  our findings show that the risk measures are very volatile in the most recent period (see also all standard deviation values in \ref{TableLower_Std}).  From our data provider, we know that these  ESG scores are not yet definitive and can be updated as analysed by \cite{Berg2021}, making the behavior in the last period subject to change. 

Looking at the lower tail market risk conditionally on ESG classes, we find that it is very close to 0.  The extreme ESG classes, $A$ and $D$, show an increased risk in the first time period; however, the magnitude is still close to 0.  Again, we find large variability in the last period, which is possibly due to the non-definitive ESG scores. Lastly,  the idiosyncratic component for lower tail risk events is very small but still seem to be highest for ESG class $A$ in 2011-2015. The variability again seems to be highest in the most recent period.

\begin{table}[H]
\centering
\footnotesize
\begin{tabular}{c|cccc|cccc|cccc}
\hline
Type of Risk &  \multicolumn{4}{c|}{Lower Tail ESG Risk}     & \multicolumn{4}{c|}{Lower Tail Market Risk}   & \multicolumn{4}{c}{Lower Tail Idiosyncratic Risk}    \\
\hline
Year         & A        & B    & C    & D    & A           & B    & C    & D    & A                  & B    & C    & D   \\
\hline
2006-2010 & 0.881 & 0.902 & 0.898 & 0.897 & 0.084 & 0.053 & 0.054 & 0.079 & 0.035 & 0.045 & 0.048 & 0.024 \\
2011-2015 & 0.911 & 0.881 & 0.937 & 0.927 & 0.035 & 0.049 & 0.028 & 0.044 & 0.054 & 0.070 & 0.036 & 0.029 \\
2016-2018 & 0.768 & 0.638 & 0.816 & 0.729 & 0.072 & 0.087 & 0.093 & 0.110 & 0.159 & 0.274 & 0.090 & 0.161 \\
 \hline
\end{tabular}
\caption{Mean values for lower tail ESG risk, Market Risk and Idiosyncratic Risk for each ESG Class $k$ in time interval $q$.}
\label{tablelower}
\end{table}

\section{Conclusion}\label{sec:Con}
Starting from an empirical analysis of real-world financial data, we notice that ESG rating classes can provide information on overall risk and tail risk. In fact, as Figure \ref{VaR} shows, especially in the time periods 2006-2010 and 2011-2015, assets with better ESG ratings seem to exhibit less tail risk.   However,  as companies tend to improve their rating throughout time, as seen in Figure \ref{distribution}, these differences tend to diminish,  especially in the interval 2016-2018, in which, however,  Refinitiv ESG scores cannot yet be considered definitive.  Based on such observations, we propose the R-vine copula ESG risk model to capture (tail) dependence and compute three new risk measures both for overall and lower tail risk. 
 
Using vine copula models allows us to capture financial times series characteristics including non-symmetry and dependence in the extremes,  as it can rely on a pair copula construction.  Especially the proposed  R-vine copula ESG risk model is able to flexibly model the dependence structures using the ESG scores.   Moreover,  as for each level an arbitrary bivariate copula can be specified, every complex dependence structure can be captured effectively and optimally.  After estimating the R-vine copula ESG risk model, we cannot only estimate all the conditional dependencies among assets as well as specify their interactions as modelled by different copulas families, but we can also introduce three ESG risk measures that capture ESG risk, market risk conditionally on the ESG class, as well as an idiosyncratic risk component. We notice that in times of calm, assets with a $D$ score show the highest individual ESG risk and tail ESG risk, however, we do not confirm this finding in times of crisis.  Assets belonging to ESG class $A$ show unexpected worse ESG risk in times of calm, possibly because of a high investment volume and popularity from investors who wrongly believe that ESG performance can make them resilient in times of crisis. 
We also find a lot of variability in the risk measures in the last years, when the ESG scores are not definitive and can still be changed.  
Companies that are not best or worst performers, leaving assets in ESG Class $B$ and $C$, often show the lowest ESG risk.  Again, this could be due to the non-popularity of assets with mediocre ESG performance.

The understanding and estimation of such dependencies and risks is of utmost importance for setting up adequate risk management and mitigation tools as well as building portfolios, ideally, also ESG diversified and resilient to crises. Current popular ESG inclusion approaches that focus on picking only assets in the highest ESG rating classes could have indeed possibly benefited in the past from better VaR values but such behavior is not clear for the most recent interval, where ESG classes are overlapping and differences are diminishing.  In fact, picking assets with the highest ESG scores does not lead to better VaR values necessarily and could result in applying too much pressure on a specific set of assets without a clear benefit. The constant trend in improving ESG scores, as shown in Figure \ref{distribution}, might be a factor behind the lack of VaR differentiation between the classes \textit{A,} \textit{B}, and \textit{C} as well as the large variability in the ESG risk in the last time interval, joint to the fact that such ESG scores are not yet definitive. Still, we notice that ESG class \textit{D} assets tend to exhibit poorer VaR values than other ESG classes and especially in times of calm, exhibit the large overall and lower tail ESG risk,  as such classes mostly include assets which have yet to disclose information needed for ESG score computation. This suggests that ESG disclosure might also have some indirect and positive effect on the company risk management. High on the agenda, the current model could be used to develop new ESG investment strategies and ESG based risk mitigation and management modelling tools.

\newpage

\appendix

\section{Notation and Computation}

\subsection{Data}\label{Information_Notation}
We introduce the mathematical indices, data sets and their notations used in the paper.

\begin{table}[H]
\centering
\small
\begin{tabular}{ l | c  }
\hline
Type of Data & Notation \\ 
 \hline
 Economics Sector  & S$ =1, \cdots, 10$\\
 Trading day & $t = 1,...,3271$\\
 Year &  $y =2006,\ldots,2018$\\
 Period  & $q=1,2,3$ \\ 
ESG class Quartiles &  $ k \in \{A, B, C, D\}$\\
\hline
ESG score of asset $j$ in year $y$ &  $ESG_{y,j}$ \\
Log return of asset $j$  on trading day $t$ &  $ Y_{t,j}$ \\
\makecell[l]{Log return of asset $j$ in period  $q$ on trading day $t$ \\  belonging to ESG class $k$} &  $ Y_{t,j,k}^{q}$ \\

S\&P 500 log return on trading day $t$  in period $q$& $I_{t,M}^{q}$\\
Market capitalization weight of asset $j$ (by 1.01.2015) & $M_{j}$ \\ 
\hline
 Mean ESG score of asset $j$and period $q$ & $\overline{ESG}^{q}_{j}$\\
 ESG class of asset $j$ and period $q$ per sector $S$ &  $K_{j}^{S,q} $\\
 ESG class weight of asset $j$  in period $q$ & $ \alpha_{j}^{q}$\\
Values of ESG class $k$ and period $q$ at trading day $t$ & $I_{t,k}^{q} $ \\
\end{tabular}
\caption{Mathematical Indices, datasets, and their notation used in the paper.}
\label{table:index}
\end{table}

As mentioned in Section \ref{data}, we worked with the mean ESG score and the corresponding ESG class of an asset $j$ for period $q$. Accordingly, we calculated the assets' ESG class weights, which were then used to calculate their corresponding ESG class values, as defined in Table \ref{table:index}.  Their computations are given as follows.

\subsection{Mean ESG score of asset $j$ and period $q$  ($\overline{ESG}^{q}_{j}$) : }\label{mean}

\begin{equation}
\overline{ESG}^{q}_{j} = \frac{1}{|P_q|} \sum_{y\in P_q}ESG_{y,j} \quad \textrm{for} \quad \forall_{j,q}, \
\end{equation}
where $P_1 = [2006, 2010]$, $P_2 = [2011, 2015]$, $P_3 = [2016, 2018]$, and $|P_q|$ denotes the number of years in $P_q$.

\subsection{Define ESG classes for each asset $j$ within sector $S$ and period $q$ using quartiles ($K_{j}^{S, q}$) : }\label{define}

Assume $j^{S,q}$ contains asset $j$ in sector $S$ and period $q$ based on their mean ESG scores $\overline{ESG}^{q}_{j}$, so that the assets are ordered non-decreasingly.  To simplify notation we remove the indices $S$ and $q$ from the asset e.g.,  $l_1 = l_1^{S,q}$. ${n_s}$ is defined as the total number of assets with in each sector $S$. Then, we define the ESG quartiles for Sector $S$ in period $q$ as follows:\\

We have $\forall_{j,S,q}$:
\begin{equation}
\begin{aligned}
& j^{S,q} = \{l_1, \cdots , l_{n_s} \} \\
& D^{S,q}= \{l_1, \cdots , l_{n_{D_s}} \},  \quad \mathrm{where} \quad  n_{D_s} = \mathrm{int} \bigg (\frac{n_s}{4} \bigg) \\
\end{aligned}
\end{equation}
\begin{equation}
\begin{aligned}
& C^{S,q}= \{l_{n_{D_s+1}}, \cdots , l_{n_{C_s}} \},  \quad \mathrm{where} \begin{cases} 
 \quad \begin{aligned} & n_{C_s} = 2  \cdot  n_{D_s}  +1 \quad
 &  \mathrm{if}
 \quad \mathrm{mod}(n_s,4)=3 \\
 & n_{C_s} = 2  \cdot  n_{D_s}  \quad
 &  \mathrm{else}
 \end{aligned}
  \end{cases}
\end{aligned}
\end{equation}

\begin{equation}
\begin{aligned}
& B^{S,q}= \{l_{n_{C_s+1}}, \cdots , l_{n_{B_s}} \},  \quad \mathrm{where} \begin{cases} 
 \quad \begin{aligned} & n_{B_s} = 3  \cdot  n_{D_s}  +1 \quad
 &  \mathrm{if}
 \quad \mathrm{mod}(n_s,4)=2,3 \\
 & n_{B_s} = 3  \cdot  n_{D_s}  \quad
 &  \mathrm{else}
 \end{aligned}
 \end{cases}
\end{aligned}
\end{equation}

\begin{equation}
 A^{S,q}= \{l_{n_{B_s+1}}, \cdots , l_{n_s} \}  
\end{equation}

\begin{equation}
K_{j}^{S,q} = 
\begin{cases} 
       & \textit{A},  \quad  \textrm{if } j \in A^{S,q}, \\
       & \textit{B},  \quad  \textrm{if } j \in B^{S,q}, \\
       & \textit{C},  \quad  \textrm{if } j \in C^{S,q}, \\
       & \textit{D},   \quad    j \in D^{S,q}. \\
\end{cases} 
\end{equation}

\subsection*{ESG class weight of asset $j$ in period $q$  :}

\begin{equation}
\alpha_{j}^{q} = \frac{M_{j}}{\sum\limits_{\substack{j’ \in [1, n] \\ j’:  K_{j'}^{S,q} = K_{j}^{S,q}}}M_{j’}} \quad \textrm{for} \quad \forall_{j,S,q}.
\end{equation}

\subsection{Values of ESG class $k$ in period $q$ at trading day $t$ :}

\begin{equation}
\bm{I}_{t,k}^{q} =  {\sum\limits_{\substack{j’ \in [1, n] \\j’:  K_{j'}^{S,q} = k}}} \alpha_{j'}^{q} \cdot Y_{t,j',k}^{q}\quad \textrm{for} \quad \forall_{q, k} \quad \textrm{and} \quad  t \in T_q,
\end{equation}

where $T_1 = [1, 1260]$, $T_2 = [1261,2517]$, $T_3 = [2518, 3271]$.

\section{Two-Step Inference for Margins} \label{IFM}

As financial data are strongly dependent on past values and not uniformly distributed on $[0,1]^d$, which is the necessary input for a copula,  a two-step inference for margins (IFM) approach is followed.  This approach as been investigated by \cite{ref:joe1}.  We follow a parametric marginal model and estimate the margins first, we then use the estimated marginal distributions to transform the data on the copula scale by defining the pseudo-copula data.  This allows us to remove the marginal time dependence by utilizing standard univariate time series models and then proceed with standardized residuals obtained from these models. 
We fit a generalized autoregressive conditional heteroskedasticity (GARCH) model with Student \emph{t} innovations to our data, allowing for time-varying volatility and volatility clustering.

\begin{table}[H]
\centering
\scalebox{0.6}{
\begin{tabular}{ l | c  }
\hline
Parameters & Notation \\ 
 \hline
The set of trading days in period $q$ & $p_q$ with $p_0 = \emptyset$,  $p_1 = \{1, \ldots, 1260\}$, $p_2 = \{1261, \ldots, 2517\}$, $p_3=\{2518, \ldots, 3271\}$\\
S\&P 500 log returns in period $q$ & $\bm{I}^{M,q}$ =$(I^M_{1+|p_{q-1}|}, \ldots, I^M_{|p_{q-1}|+|p_q|})^\top  \in \mathbb{R}^{|p_q|}$\\
Matrix of log returns $Y_{t,j}$ in sector and period $q$ for $T \in t_q$ & $Y^{q} = [\bm{Y}_{1}, \ldots, \bm{Y}_{n}]\in \mathbb{R}^{|p_q|\times n} $, where $ \bm{Y}_{j'} =(Y_{1+|p_{q-1}|,j'}, \ldots, Y_{|p_{q-1}|+|p_q|,j'})^{\top}$, $j’ \in [1, n]$ \\
Data matrix for  in period $q$ &  $X^{q} = [ Y^{q}$,$\bm{I}^{q}, \bm{I}_{A}^{q}, \bm{I}_{B}^{q},  \bm{I}_{C}^{q}, \bm{I}_{D}^{q}] \in \mathbb{R}^{|p_q|\times P}$\\
Columns of the data matrix $X^{q} $ & $d=1,\ldots,P$ \\
Rows of the data matrix $X^{q} $ & $i=1,\ldots,|p_q|$ \\
Column $d$ of the data matrix $X^{q} $ & $\bm{X}_d^{q} $ =$(X_{1,d}^{q}, \ldots, X_{|p_q|,d}^{q})^\top \in \mathbb{R}^{|p_q|}$\\
Conditional variance vector of $\bm{X}_{t,d}^{q} $ on trading day $t$& $(\sigma_{d,t}^{q})^2$\\
Estimated degree of freedom for $\bm{X}_{t,d}^{q} $  &$\hat{\nu}_{d}$  \\
Estimated covariance for $\bm{X}_d^{q} $  on trading day $t$ & $\hat{\sigma}_{d,t} ^{2}$ \\
Estimated distribution function of the innovation distribution for time series $\bm{X}_{t,d}^{q} $ & $\hat{F}^{q}_d(\; \cdot \;     ;  \hat{\nu}_{d})$ \\
Estimated u-data for an observation $X_{t,d}^{q}$ in period $q$  & $\hat{u}_{t,d}^{q}$ where $t \in T_q$ \\
\end{tabular}}
 \caption{Notation of the GARCH and Copula model.}
\label{table:GARCH-data}
\end{table}

As an input of a R-vine model in  period $q$, we have a data matrix $X^{q}$ defined in Table \ref{table:GARCH-data}. 

In  period $q$, we fit a GARCH(1,1) model with appropriate error distribution for a marginal time series, $\bm{X}_d^{q}$, and estimate the parameters of the following model: 

\begin{equation}
\varepsilon_{d,t}^{q} =  \sigma_{t,d}^{q} \cdot z_{t} \quad  \quad  \quad (\sigma_{d,t}^{q})^2 = \gamma_0 + \gamma_1 \cdot (\varepsilon_{t-1,d}^{q})^2 + \beta_1 \cdot (\sigma_{t-1,d}^{q})^2
\end{equation}

where $(z_t)_{t>1}$ is a sequence of normal random independent and identically distributed random variables satisfying the standard  assumptions $E[z_t]=0$ and $var[z_t]=1$ and follows a Student's \emph{t} distribution.  Then using the cumulative distribution function of the standardized Student's \emph{t} distribution,  we determine the  \emph{pseudo-copula data} with the probability integral transformation (PIT), i.e.

 \begin{equation}
\hat{u}_{t,d}^{q} := \hat{F}_d \left(\frac{ X_{t,d}^{q}}{\hat{\sigma}_{t,d}} ; \hat{\nu}_{d}\right). 
 \end{equation}
 
Following this two-step approach allows us to convert data to the copula scale,   which serves as the input for our R-vine copula ESG risk model.

\section{Choosing the Bivariate Copula Families and Estimating the Copula Parameters}\label{Familyselection}

Then, from the set of bivariate copula families in Table \ref{DifCop},  the optimal pair copula families for the variable pairs are selected using the Akaike Information Criterion (AIC) \citep{ref:akaike}, which has been shown to have good copula family selection properties and high accuracy \citep{Brechmann2010}.  Since each pair copula family can be specified sequentially and independently at the same tree level, an alternative selection criteria like the BIC would not be needed to induce sparsity at this step.  In the next step, we compute the pseudo copula data as defined in \ref{IFM} and also more detailed in \cite{ref:czado1} for the second tree level $T_2$, using the estimated pair copulas in the first tree level $T_1$. This input data is similarly  used to select the second tree level $T_2$ of the vine and corresponding pair copula families with their parameters. The approach continues sequentially up to the last tree level. Following this method, we select the pair copula families, fit the five vine trees, and estimate the parameter values. As we choose a pair copula family corresponding to each edge separately, the parameter estimation step is at most a two-dimensional optimization, which is computationally efficient. Moreover, the performance of this method is satisfactory compared to the full log likelihood method, which could be computationally intractable in high dimensions  \citep{Haff2012}. More details about R-vines and vine trees are given by \cite{ref:kura, ref:kurojoe, ref:joe2014}; and \cite{ref:czado1}.

\begin{table}[hbt]
\centering
\footnotesize
\begin{tabular}{l|ccccccc|cccc} 
\hline
           & \multicolumn{7}{c|}{Itau\textsuperscript{1} Copulas}  & \multicolumn{3}{c}{ BB Copulas}        &            \\
                \hline

  Properties                     & \emph{t}             &  \emph{F}                   & \emph{N}                 &  \emph{C}                   &\emph{J}                     &  \emph{G}                    &  \textbf{\emph{I}} &  \emph{BB1} & \emph{BB7}  & \emph{BB8} &\\
                       \hline
                       \hline 
Positive Dependence    & \checkmark & \checkmark & \checkmark & \checkmark & \checkmark & \checkmark &  -&  \checkmark &  \checkmark &  \checkmark & \\
Negative Dependence    & \checkmark & \checkmark & \checkmark & -                         & -                         & -                         & -&  -&- & -& \\
Tail Asymmetry          & -                         & -                         & -                         & \checkmark & \checkmark & \checkmark &  -&  \checkmark & \checkmark &  \checkmark & \\
Lower Tail Dependence  & \checkmark & -                         & -                         & \checkmark & -                         & -                         &  -&   \checkmark &  \checkmark & - & \\
Upper Tail Dependence & \checkmark & -                         & -                         & -                         & \checkmark & \checkmark & -&  \checkmark & \checkmark &  \checkmark & \\
\hline
\end{tabular}
\caption{Parametric copula families and their properties without rotations and reflections. Notation of copula families:t = Student’s t,  F = Frank, N = Gaussian,  C = Clayton, J = Joe, G = Gumbel,  \textbf{\emph{I}}  = Independence,  BB1 = Clayton- Gumbel,  BB7 = Joe-Clayton,  BB8= Extended Joe.}
\textsuperscript{1}\footnotesize{Copula families  for which the parameter estimation by Kendall’s $\tau$ inversion is available without rotations.}
\label{DifCop}
\end{table}

To extend the range of dependence,  counterclockwise rotations and reflections of the copula density are included.  This allows the Gumbel, Clayton, Joe, BB1, BB7 and BB8 to  also accommodate negative dependence ($\tau < 0)$.  For more details, we refer to Chapter 3 of  \cite{ref:czado1}. 

\section{Model Fit} 

\subsection{mBIC of the R-vine copula ESG risk model  with three specifications} \label{Modelfit}

Comparison of the \emph{itau} and \emph{parametric}  and \emph{Gaussian} R-vine models.
\begin{table}[H]
\centering
\begin{tabular}{l|l|llll}
\hline
Model & Year      & nobs & logLik    & npars & mBIC              \\
\hline
itau  & 2006-2010 & 1260 & 133415.85 & 2300  & \textbf{-241681.52} \\
par   & 2006-2011 & 1260 & 133730.31 & 2427  & -241396.61 \\
gaus  & 2006-2012 & 1260 & 123607.92 & 1680  & -226242.07 \\
\hline
itau  & 2011-2015 & 1257 & 132119.36 & 1975  & \textbf{-241564.51 }\\
par   & 2011-2016 & 1257 & 132430.2  & 2118  & -241174.01 \\
gaus  & 2011-2017 & 1257 & 123696.65 & 1680  & -226423.54 \\
\hline
itau  & 2016-2018 & 754  & 52671.04  & 1843  & \textbf{-84884.99} \\
par   & 2016-2019 & 754  & 52882.97  & 1944  & -84646.93 \\
gaus  & 2016-2020 & 754  & 48514.13  & 1680  & -76917.14 \\
\hline
\end{tabular}
\caption{Model fit for each model at different time interval $q$- decision measure mBIC \citep{Nagler2019}.  Additional information are available upon author request. }
\end{table}

\subsection{Copula Families Estimated from 7 different copula families and their rotations}\label{Families}
\begin{table}[H]
\footnotesize
\centering
\begin{tabular}{l|ccc|ccc|ccc}
             \hline
       Copula Family \& Rotation        &\multicolumn{3}{c|}{Itau Copula Families}         & \multicolumn{3}{c|}{Parametric Copula Families}         & \multicolumn{3}{c}{Gaussian Copula Families}   \\
              \hline
 Year          & 2006-2010 & 2011-2015 & 2016-2018 & 2006-2010 & 2011-2015 & 2016-2018 & 2006-2010 & 2011-2015 & 2016-2018 \\
       \hline
Studentst    & 325       & 319       & 289       & 287       & 296       & 234       & 0         & 0         & 0         \\
Clayton      & 0         & 0         & 1         & 0         & 0         & 1         & 0         & 0         & 0         \\
Frank        & 8         & 9         & 15        & 3         & 5         & 10        & 0         & 0         & 0         \\
Gaussian     & 2         & 1         & 20        & 2         & 1         & 16        & 338       & 338       & 338       \\
Gumbel       & 0         & 5         & 2         & 0         & 0         & 2         & 0         & 0         & 0         \\
Independence & 0         & 0         & 4         & 0         & 0         & 4         & 0         & 0         & 0         \\
Gumbel 180°    & 3         & 4         & 7         & 0         & 2         & 1         & 0         & 0         & 0         \\
BB1          & 0         & 0         & 0         & 0         & 2         & 0         & 0         & 0         & 0         \\
BB7          & 0         & 0         & 0         & 1         & 0         & 0         & 0         & 0         & 0         \\
BB8          & 0         & 0         & 0         & 15        & 26        & 12        & 0         & 0         & 0         \\
BB1 90°   & 0         & 0         & 0         & 3         & 0         & 0         & 0         & 0         & 0         \\
BB7 90°      & 0         & 0         & 0         & 0         & 0         & 1         & 0         & 0         & 0         \\
BB8 90 °     & 0         & 0         & 0         & 2         & 0         & 1         & 0         & 0         & 0         \\
BB1 180°     & 0         & 0         & 0         & 2         & 3         & 1         & 0         & 0         & 0         \\
BB7 180°    & 0         & 0         & 0         & 3         & 1         & 0         & 0         & 0         & 0         \\
BB8 180°     & 0         & 0         & 0         & 16        & 12        & 9         & 0         & 0         & 0         \\
BB1 270°     & 0         & 0         & 0         & 3         & 0         & 1         & 0         & 0         & 0         \\
BB7 270°     & 0         & 0         & 0         & 1         & 2         & 3         & 0         & 0         & 0         \\
BB8 270°     & 0         & 0         & 0         & 2         & 0         & 1         & 0         & 0         & 0        \\
             \hline
\end{tabular}
\caption{Bivariate copula families and independence copula fitted for Tree 1 ($T_1$).  Only copula families which are chosen at least once are presented.   Additional vine trees are available upon author request. }
\end{table}

\section{Additional Information on Risk Measures}
\subsection{Standard Deviation of Risk Measures}

\begin{table}[h]\label{TableOverall_Std}
\centering
\footnotesize
\begin{tabular}{c|cccc|cccc|cccc}
\hline
Type of Risk &  \multicolumn{4}{c|}{ESG Risk}     & \multicolumn{4}{c|}{Market Risk}   & \multicolumn{4}{c}{Idiosyncratic Risk}    \\
\hline
Year         & A        & B    & C    & D    & A           & B    & C    & D    & A                  & B    & C    & D   \\
\hline
2006-2010 & 0.124 & 0.087 & 0.086 & 0.086 & 0.102 & 0.056 & 0.058 & 0.074 & 0.088 & 0.080 & 0.075 & 0.074 \\
2011-2015 & 0.093 & 0.104 & 0.090 & 0.081 & 0.045 & 0.041 & 0.039 & 0.040 & 0.091 & 0.086 & 0.078 & 0.081 \\
2016-2018 & 0.178 & 0.123 & 0.125 & 0.124 & 0.036 & 0.040 & 0.037 & 0.042 & 0.163 & 0.118 & 0.120 & 0.114 \\
\hline
\end{tabular}
\caption{Standard Deviation values for overall ESG Risk, Market Risk and Idiosyncratic Risk for each ESG Class $k$ in time interval $q$}
\end{table}

\begin{table}[h]\label{TableLower_Std}
\centering
\footnotesize
\begin{tabular}{c|cccc|cccc|cccc}
\hline
Type of Risk &  \multicolumn{4}{c|}{ESG Risk}     & \multicolumn{4}{c|}{Market Risk}   & \multicolumn{4}{c}{Idiosyncratic Risk}    \\
\hline
Year         & A        & B    & C    & D    & A           & B    & C    & D    & A                  & B    & C    & D   \\
\hline
2006-2010 & 0.147 & 0.135 & 0.186 & 0.293 & 0.116 & 0.073 & 0.077 & 0.228 & 0.105 & 0.119 & 0.180 & 0.170 \\
2011-2015 & 0.295 & 0.207 & 0.215 & 0.166 & 0.020 & 0.026 & 0.018 & 0.016 & 0.293 & 0.207 & 0.215 & 0.167 \\
2016-2018 & 0.330 & 0.335 & 0.348 & 0.303 & 0.030 & 0.065 & 0.123 & 0.122 & 0.330 & 0.331 & 0.342 & 0.278 \\
 \hline
\end{tabular}
\caption{Standard Deviation values for lower tail ESG risk, Market Risk and Idiosyncratic Risk for each ESG Class $k$ in time interval $q$}
\end{table}

\newpage
\bibliography{Paper_1_extracted.bib}

\begin{thebibliography}{92}
\expandafter\ifx\csname natexlab\endcsname\relax\def\natexlab#1{#1}\fi
\providecommand{\url}[1]{\texttt{#1}}
\providecommand{\href}[2]{#2}
\providecommand{\path}[1]{#1}
\providecommand{\DOIprefix}{doi:}
\providecommand{\ArXivprefix}{arXiv:}
\providecommand{\URLprefix}{URL: }
\providecommand{\Pubmedprefix}{pmid:}
\providecommand{\doi}[1]{\href{http://dx.doi.org/#1}{\path{#1}}}
\providecommand{\Pubmed}[1]{\href{pmid:#1}{\path{#1}}}
\providecommand{\bibinfo}[2]{#2}
\ifx\xfnm\relax \def\xfnm[#1]{\unskip,\space#1}\fi
\bibitem[{Aas et~al.(2009)Aas, Czado, Frigessi \& Bakken}]{ref:aas}
\bibinfo{author}{Aas, K.}, \bibinfo{author}{Czado, C.},
  \bibinfo{author}{Frigessi, A.}, \& \bibinfo{author}{Bakken, H.}
  (\bibinfo{year}{2009}).
\newblock \bibinfo{title}{{Pair-copula constructions of multiple dependence}}.
\newblock {\it \bibinfo{journal}{Insurance: Mathematics and Economics}\/},
  {\it \bibinfo{volume}{44}\/}, \bibinfo{pages}{182--198}. \URLprefix
  \url{http://dx.doi.org/10.1016/j.insmatheco.2007.02.001}.
  \DOIprefix\doi{10.1016/j.insmatheco.2007.02.001}.
\bibitem[{Abakah et~al.(2021)Abakah, {Addo Jr}, Gil-Alana \&
  Tiwari}]{Abakah2021}
\bibinfo{author}{Abakah, E. J.~A.}, \bibinfo{author}{{Addo Jr}, E.},
  \bibinfo{author}{Gil-Alana, L.~A.}, \& \bibinfo{author}{Tiwari, A.~K.}
  (\bibinfo{year}{2021}).
\newblock \bibinfo{title}{{Re-examination of international bond market
  dependence: Evidence from a pair copula approach}}.
\newblock {\it \bibinfo{journal}{International Review of Financial
  Analysis}\/},  {\it \bibinfo{volume}{74}\/}, \bibinfo{pages}{101678}.
\bibitem[{Achim \& Borlea(2015)}]{ref:achim}
\bibinfo{author}{Achim, M.-V.}, \& \bibinfo{author}{Borlea, S.~N.}
  (\bibinfo{year}{2015}).
\newblock \bibinfo{title}{Developing of esg score to assess the non-financial
  performances in romanian companies}.
\newblock {\it \bibinfo{journal}{Procedia Economics and Finance}\/},  {\it
  \bibinfo{volume}{32}\/}, \bibinfo{pages}{1209--1224}.
  \DOIprefix\doi{https://doi.org/10.1016/S2212-5671(15)01499-9}.
\newblock \bibinfo{note}{Emerging Markets Queries in Finance and Business 2014,
  EMQFB 2014, 24-25 October 2014, Bucharest, Romania}.
\bibitem[{Akaike(1998)}]{ref:akaike}
\bibinfo{author}{Akaike, H.} (\bibinfo{year}{1998}).
\newblock \bibinfo{title}{{Information theory and an extension of the maximum
  likelihood principle}}.
\newblock In {\it \bibinfo{booktitle}{Parzen E., Tanabe K., Kitagawa G. (eds)
  Selected Papers of Hirotugu Akaike. Springer Series in Statistics
  (Perspectives in Statistics)}\/}.
\newblock \bibinfo{publisher}{Springer}.
\newblock \DOIprefix\doi{https://doi.org/10.1007/978-1-4612-1694-0_15}.
\bibitem[{Albuquerque et~al.(2020)Albuquerque, Koskinen, Yang \&
  Zhang}]{Albuquerque2020}
\bibinfo{author}{Albuquerque, R.}, \bibinfo{author}{Koskinen, Y.},
  \bibinfo{author}{Yang, S.}, \& \bibinfo{author}{Zhang, C.}
  (\bibinfo{year}{2020}).
\newblock \bibinfo{title}{{Resiliency of environmental and social stocks: An
  analysis of the exogenous COVID-19 market crash}}.
\newblock {\it \bibinfo{journal}{Review of Corporate Finance Studies}\/},  {\it
  \bibinfo{volume}{9}\/}, \bibinfo{pages}{593--621}.
\bibitem[{Ane \& Kharoubi(2003)}]{ref:ane}
\bibinfo{author}{Ane, T.}, \& \bibinfo{author}{Kharoubi, C.}
  (\bibinfo{year}{2003}).
\newblock \bibinfo{title}{{Dependence structure and risk measure}}.
\newblock {\it \bibinfo{journal}{Journal of Business}\/},  {\it
  \bibinfo{volume}{76}\/}, \bibinfo{pages}{411--438}.
\bibitem[{{Ashwin Kumar} et~al.(2016){Ashwin Kumar}, Smith, Badis, Wang,
  Ambrosy \& Tavares}]{Kumar2016}
\bibinfo{author}{{Ashwin Kumar}, N.~C.}, \bibinfo{author}{Smith, C.},
  \bibinfo{author}{Badis, L.}, \bibinfo{author}{Wang, N.},
  \bibinfo{author}{Ambrosy, P.}, \& \bibinfo{author}{Tavares, R.}
  (\bibinfo{year}{2016}).
\newblock \bibinfo{title}{{ESG factors and risk-adjusted performance: a new
  quantitative model}}.
\newblock {\it \bibinfo{journal}{Journal of Sustainable Finance {\&}
  Investment}\/},  {\it \bibinfo{volume}{6}\/}, \bibinfo{pages}{292--300}.
\bibitem[{Auer \& Schuhmacher(2016)}]{ref:auer}
\bibinfo{author}{Auer, B.~R.}, \& \bibinfo{author}{Schuhmacher, F.}
  (\bibinfo{year}{2016}).
\newblock \bibinfo{title}{{Do socially (ir) responsible investments pay? New
  evidence from international ESG data}}.
\newblock {\it \bibinfo{journal}{The Quarterly Review of Economics and
  Finance}\/},  {\it \bibinfo{volume}{59}\/}, \bibinfo{pages}{51--62}.
\bibitem[{Bae et~al.(2019)Bae, {El Ghoul}, Guedhami, Kwok \& Zheng}]{Bae2019}
\bibinfo{author}{Bae, K.-H.}, \bibinfo{author}{{El Ghoul}, S.},
  \bibinfo{author}{Guedhami, O.}, \bibinfo{author}{Kwok, C. C.~Y.}, \&
  \bibinfo{author}{Zheng, Y.} (\bibinfo{year}{2019}).
\newblock \bibinfo{title}{{Does corporate social responsibility reduce the
  costs of high leverage? Evidence from capital structure and product market
  interactions}}.
\newblock {\it \bibinfo{journal}{Journal of Banking {\&} Finance}\/},  {\it
  \bibinfo{volume}{100}\/}, \bibinfo{pages}{135--150}.
\bibitem[{Becchetti et~al.(2015)Becchetti, Ciciretti \& Hasan}]{Becchetti2015}
\bibinfo{author}{Becchetti, L.}, \bibinfo{author}{Ciciretti, R.}, \&
  \bibinfo{author}{Hasan, I.} (\bibinfo{year}{2015}).
\newblock \bibinfo{title}{{Corporate social responsibility, stakeholder risk,
  and idiosyncratic volatility}}.
\newblock {\it \bibinfo{journal}{Journal of Corporate Finance}\/},  {\it
  \bibinfo{volume}{35}\/}, \bibinfo{pages}{297--309}.
\bibitem[{Bedford \& Cooke(2001)}]{ref:bedford2}
\bibinfo{author}{Bedford, T.}, \& \bibinfo{author}{Cooke, R.~M.}
  (\bibinfo{year}{2001}).
\newblock \bibinfo{title}{{Probability density decomposition for conditionally
  dependent random variables modeled by vines}}.
\newblock {\it \bibinfo{journal}{Annals of Mathematics and Artificial
  intelligence}\/},  {\it \bibinfo{volume}{32}\/}, \bibinfo{pages}{245--268}.
\bibitem[{Bedford \& Cooke(2002)}]{ref:bedford1}
\bibinfo{author}{Bedford, T.}, \& \bibinfo{author}{Cooke, R.~M.}
  (\bibinfo{year}{2002}).
\newblock \bibinfo{title}{{Vines : A New Graphical Model for Dependent Random
  Variables}}.
\newblock {\it \bibinfo{journal}{Annals of Statistics}\/},  {\it
  \bibinfo{volume}{30}\/}, \bibinfo{pages}{1031--1068}.
\bibitem[{BenSa{\"{i}}da(2018)}]{Ben2018}
\bibinfo{author}{BenSa{\"{i}}da, A.} (\bibinfo{year}{2018}).
\newblock \bibinfo{title}{{The contagion effect in European sovereign debt
  markets: A regime-switching vine copula approach}}.
\newblock {\it \bibinfo{journal}{International Review of Financial
  Analysis}\/},  {\it \bibinfo{volume}{58}\/}, \bibinfo{pages}{153--165}.
\bibitem[{Berg \& Lange(2020)}]{ref:berg}
\bibinfo{author}{Berg, E.}, \& \bibinfo{author}{Lange, K.~W.}
  (\bibinfo{year}{2020}).
\newblock {\it \bibinfo{title}{{Enhancing ESG-Risk Modelling-A study of the
  dependence structure of sustainable investing}}\/}.
\newblock Master's thesis KTH Royal Institute of Technology.
\newblock \URLprefix \url{http://hdl.handle.net/2445/169669}
  \bibinfo{note}{accessed on 03-11-2021}.
\bibitem[{Berg et~al.(2021)Berg, Fabisik \& Sautner}]{Berg2021}
\bibinfo{author}{Berg, F.}, \bibinfo{author}{Fabisik, K.}, \&
  \bibinfo{author}{Sautner, Z.} (\bibinfo{year}{2021}).
\newblock \bibinfo{title}{{Is History Repeating Itself? The (Un) predictable
  Past of ESG Ratings}}.
\newblock {\it \bibinfo{journal}{European Corporate Governance Institute –
  Finance Working Paper}\/},  {\it \bibinfo{volume}{708/2020}\/}.
  \DOIprefix\doi{10.2139/ssrn.3722087}.
\newblock \bibinfo{note}{Accessed on 03-11-2021}.
\bibitem[{Berg et~al.(2019)Berg, Koelbel \& Rigobon}]{Berg2019}
\bibinfo{author}{Berg, F.}, \bibinfo{author}{Koelbel, J.~F.}, \&
  \bibinfo{author}{Rigobon, R.} (\bibinfo{year}{2019}).
\newblock \bibinfo{title}{{Aggregate Confusion: The Divergence of ESG
  Ratings}}.
\newblock {\it \bibinfo{journal}{SSRN Electronic Journal}\/}, .
  \DOIprefix\doi{10.2139/ssrn.3438533}.
\newblock \bibinfo{note}{Accessed on 03-11-2021}.
\bibitem[{Bhattacharya \& Sharma(2019)}]{ref:bhatta}
\bibinfo{author}{Bhattacharya, S.}, \& \bibinfo{author}{Sharma, D.}
  (\bibinfo{year}{2019}).
\newblock \bibinfo{title}{{Do environment, social and governance performance
  impact credit ratings: a study from India}}.
\newblock {\it \bibinfo{journal}{International Journal of Ethics and
  Systems}\/},  {\it \bibinfo{volume}{35}\/}, \bibinfo{pages}{466--484}.
\bibitem[{Bhatti \& Nguyen(2012)}]{Bhatti2012}
\bibinfo{author}{Bhatti, M.~I.}, \& \bibinfo{author}{Nguyen, C.~C.}
  (\bibinfo{year}{2012}).
\newblock \bibinfo{title}{{Diversification evidence from international equity
  markets using extreme values and stochastic copulas}}.
\newblock {\it \bibinfo{journal}{Journal of International Financial Markets,
  Institutions and Money}\/},  {\it \bibinfo{volume}{22}\/},
  \bibinfo{pages}{622--646}.
\bibitem[{Billio et~al.(2021)Billio, Costola, Hristova, Latino \&
  Pelizzon}]{Billio2021}
\bibinfo{author}{Billio, M.}, \bibinfo{author}{Costola, M.},
  \bibinfo{author}{Hristova, I.}, \bibinfo{author}{Latino, C.}, \&
  \bibinfo{author}{Pelizzon, L.} (\bibinfo{year}{2021}).
\newblock \bibinfo{title}{{Inside the ESG Ratings:(Dis) agreement and
  performance}}.
\newblock {\it \bibinfo{journal}{Corporate Social Responsibility and
  Environmental Management}\/},  {\it \bibinfo{volume}{28}\/},
  \bibinfo{pages}{1426--1445}.
\bibitem[{Bouslah et~al.(2018)Bouslah, Kryzanowski \& M'Zali}]{Bouslah2018}
\bibinfo{author}{Bouslah, K.}, \bibinfo{author}{Kryzanowski, L.}, \&
  \bibinfo{author}{M'Zali, B.} (\bibinfo{year}{2018}).
\newblock \bibinfo{title}{{Social performance and firm risk: Impact of the
  financial crisis}}.
\newblock {\it \bibinfo{journal}{Journal of Business Ethics}\/},  {\it
  \bibinfo{volume}{149}\/}, \bibinfo{pages}{643--669}.
\bibitem[{Brechmann(2010)}]{Brechmann2010}
\bibinfo{author}{Brechmann, E.} (\bibinfo{year}{2010}).
\newblock {\it \bibinfo{title}{{Truncated and simplified regular vines and
  their applications}}\/}.
\newblock Master's thesis Technical University of Munich.
\newblock \URLprefix \url{https://mediatum.ub.tum.de/doc/1079285/1079285.pdf}
  \bibinfo{note}{accessed on 03-11-2021}.
\bibitem[{Brechmann \& Czado(2013)}]{ref:brechmann}
\bibinfo{author}{Brechmann, E.~C.}, \& \bibinfo{author}{Czado, C.}
  (\bibinfo{year}{2013}).
\newblock \bibinfo{title}{{Risk management with high-dimensional vine copulas:
  An analysis of the Euro Stoxx 50}}.
\newblock {\it \bibinfo{journal}{Statistics and Risk Modeling}\/},  {\it
  \bibinfo{volume}{30}\/}, \bibinfo{pages}{307--342}.
\bibitem[{Breedt et~al.(2019)Breedt, Ciliberti, Gualdi \& Seager}]{ref:breedt}
\bibinfo{author}{Breedt, A.}, \bibinfo{author}{Ciliberti, S.},
  \bibinfo{author}{Gualdi, S.}, \& \bibinfo{author}{Seager, P.}
  (\bibinfo{year}{2019}).
\newblock \bibinfo{title}{{Is ESG an Equity Factor or Just an Investment
  Guide?}}
\newblock {\it \bibinfo{journal}{Journal of Investing}\/},  {\it
  \bibinfo{volume}{28}\/}, \bibinfo{pages}{32--42}.
\bibitem[{Breuer et~al.(2018)Breuer, M{\"{u}}ller, Rosenbach \&
  Salzmann}]{Breuer2018}
\bibinfo{author}{Breuer, W.}, \bibinfo{author}{M{\"{u}}ller, T.},
  \bibinfo{author}{Rosenbach, D.}, \& \bibinfo{author}{Salzmann, A.}
  (\bibinfo{year}{2018}).
\newblock \bibinfo{title}{{Corporate social responsibility, investor
  protection, and cost of equity: A cross-country comparison}}.
\newblock {\it \bibinfo{journal}{Journal of Banking {\&} Finance}\/},  {\it
  \bibinfo{volume}{96}\/}, \bibinfo{pages}{34--55}.
\bibitem[{Broadstock et~al.(2021)Broadstock, Chan, Cheng \&
  Wang}]{Broadstock2021}
\bibinfo{author}{Broadstock, D.~C.}, \bibinfo{author}{Chan, K.},
  \bibinfo{author}{Cheng, L. T.~W.}, \& \bibinfo{author}{Wang, X.}
  (\bibinfo{year}{2021}).
\newblock \bibinfo{title}{{The role of ESG performance during times of
  financial crisis: Evidence from COVID-19 in China}}.
\newblock {\it \bibinfo{journal}{Finance Research Letters}\/},  {\it
  \bibinfo{volume}{38}\/}, \bibinfo{pages}{101716}.
\bibitem[{Campbell et~al.(2001)Campbell, Lettau, Malkiel \& Xu}]{Campbell2001}
\bibinfo{author}{Campbell, J.~Y.}, \bibinfo{author}{Lettau, M.},
  \bibinfo{author}{Malkiel, B.~G.}, \& \bibinfo{author}{Xu, Y.}
  (\bibinfo{year}{2001}).
\newblock \bibinfo{title}{{Have individual stocks become more volatile? An
  empirical exploration of idiosyncratic risk}}.
\newblock {\it \bibinfo{journal}{Journal of Finance}\/},  {\it
  \bibinfo{volume}{56}\/}, \bibinfo{pages}{1--43}.
\bibitem[{Capelle-Blancard et~al.(2019)Capelle-Blancard, Crifo, Diaye,
  Oueghlissi \& Scholtens}]{Capelle2019}
\bibinfo{author}{Capelle-Blancard, G.}, \bibinfo{author}{Crifo, P.},
  \bibinfo{author}{Diaye, M.-A.}, \bibinfo{author}{Oueghlissi, R.}, \&
  \bibinfo{author}{Scholtens, B.} (\bibinfo{year}{2019}).
\newblock \bibinfo{title}{{Sovereign bond yield spreads and sustainability: An
  empirical analysis of OECD countries}}.
\newblock {\it \bibinfo{journal}{Journal of Banking {\&} Finance}\/},  {\it
  \bibinfo{volume}{98}\/}, \bibinfo{pages}{156--169}.
\bibitem[{Chan \& Walter(2014)}]{ref:chan}
\bibinfo{author}{Chan, P.~T.}, \& \bibinfo{author}{Walter, T.}
  (\bibinfo{year}{2014}).
\newblock \bibinfo{title}{{Investment performance of
  “environmentally-friendly” firms and their initial public offers and
  seasoned equity offers}}.
\newblock {\it \bibinfo{journal}{Journal of Banking {\&} Finance}\/},  {\it
  \bibinfo{volume}{44}\/}, \bibinfo{pages}{177--188}.
\bibitem[{Consolandi et~al.(2020)Consolandi, Eccles \& Gabbi}]{Consolandi2020}
\bibinfo{author}{Consolandi, C.}, \bibinfo{author}{Eccles, R.~G.}, \&
  \bibinfo{author}{Gabbi, G.} (\bibinfo{year}{2020}).
\newblock \bibinfo{title}{{How material is a material issue? Stock returns and
  the financial relevance and financial intensity of ESG materiality}}.
\newblock {\it \bibinfo{journal}{Journal of Sustainable Finance {\&}
  Investment}\/},  {\it \bibinfo{volume}{0}\/}, \bibinfo{pages}{1--24}.
\bibitem[{Cont(2001)}]{Cont2001}
\bibinfo{author}{Cont, R.} (\bibinfo{year}{2001}).
\newblock \bibinfo{title}{{Empirical properties of asset returns: stylized
  facts and statistical issues}}.
\newblock {\it \bibinfo{journal}{Journal of Quantitative Finance}\/},  {\it
  \bibinfo{volume}{1}\/}, \bibinfo{pages}{223--236}.
\bibitem[{Cornell(2021)}]{Cornell2021}
\bibinfo{author}{Cornell, B.} (\bibinfo{year}{2021}).
\newblock \bibinfo{title}{{ESG preferences, risk and return}}.
\newblock {\it \bibinfo{journal}{European Financial Management}\/},  {\it
  \bibinfo{volume}{27}\/}, \bibinfo{pages}{12--19}.
\bibitem[{Czado(2019)}]{ref:czado1}
\bibinfo{author}{Czado, C.} (\bibinfo{year}{2019}).
\newblock \bibinfo{title}{Analyzing dependent data with vine copulas: A
  practical guide with r.}
\newblock In {\it \bibinfo{booktitle}{Lecture Notes in Statistics}\/}.
\newblock \bibinfo{publisher}{Springer}.
\newblock \DOIprefix\doi{10.1007/978-3-030-13785-4}.
\bibitem[{Czado \& Nagler(2022)}]{CzadoNagler2022}
\bibinfo{author}{Czado, C.}, \& \bibinfo{author}{Nagler, T.}
  (\bibinfo{year}{2022}).
\newblock \bibinfo{title}{Vine copula based modeling}.
\newblock {\it \bibinfo{journal}{Annual Review of Statistics and Its
  Application}\/},  {\it \bibinfo{volume}{9}\/}.
  \DOIprefix\doi{10.1146/annurev-statistics-040220-101153}.
\bibitem[{De \& Clayman(2015)}]{De2015}
\bibinfo{author}{De, I.}, \& \bibinfo{author}{Clayman, M.~R.}
  (\bibinfo{year}{2015}).
\newblock \bibinfo{title}{{The benefits of socially responsible investing: An
  active manager's perspective}}.
\newblock {\it \bibinfo{journal}{Journal of Investing}\/},  {\it
  \bibinfo{volume}{24}\/}, \bibinfo{pages}{49--72}.
\bibitem[{Demers et~al.(2021)Demers, Hendrikse, Joos \& Lev}]{Demers2021}
\bibinfo{author}{Demers, E.}, \bibinfo{author}{Hendrikse, J.},
  \bibinfo{author}{Joos, P.}, \& \bibinfo{author}{Lev, B.}
  (\bibinfo{year}{2021}).
\newblock \bibinfo{title}{{ESG did not immunize stocks during the COVID‐19
  crisis, but investments in intangible assets did}}.
\newblock {\it \bibinfo{journal}{Journal of Business Finance {\&}
  Accounting}\/},  {\it \bibinfo{volume}{48}\/}, \bibinfo{pages}{433--462}.
\bibitem[{Diemont et~al.(2016)Diemont, Moore \& Soppe}]{Dietmont2016}
\bibinfo{author}{Diemont, D.}, \bibinfo{author}{Moore, K.}, \&
  \bibinfo{author}{Soppe, A.} (\bibinfo{year}{2016}).
\newblock \bibinfo{title}{{The downside of being responsible: Corporate social
  responsibility and tail risk}}.
\newblock {\it \bibinfo{journal}{Journal of Business Ethics}\/},  {\it
  \bibinfo{volume}{137}\/}, \bibinfo{pages}{213--229}.
\bibitem[{Dorfleitner et~al.(2016)Dorfleitner, Halbritter \&
  Nguyen}]{Dorfleitner2016}
\bibinfo{author}{Dorfleitner, G.}, \bibinfo{author}{Halbritter, G.}, \&
  \bibinfo{author}{Nguyen, M.} (\bibinfo{year}{2016}).
\newblock \bibinfo{title}{{The risk of social responsibility–is it
  systematic?}}
\newblock {\it \bibinfo{journal}{Journal of Sustainable Finance {\&}
  Investment}\/},  {\it \bibinfo{volume}{6}\/}, \bibinfo{pages}{1--14}.
\bibitem[{Eccles \& Klimenko(2019)}]{Eccles2019}
\bibinfo{author}{Eccles, R.~G.}, \& \bibinfo{author}{Klimenko, S.}
  (\bibinfo{year}{2019}).
\newblock \bibinfo{title}{{The investor revolution}}.
\newblock {\it \bibinfo{journal}{Harvard Business Review}\/},  {\it
  \bibinfo{volume}{97}\/}, \bibinfo{pages}{106--116}.
\bibitem[{Embrechts et~al.(2001)Embrechts, Lindskog \& McNeil}]{Embrecht2001}
\bibinfo{author}{Embrechts, P.}, \bibinfo{author}{Lindskog, F.}, \&
  \bibinfo{author}{McNeil, A.} (\bibinfo{year}{2001}).
\newblock {\it \bibinfo{title}{{Modelling dependence with copulas}}\/}.
\newblock \bibinfo{type}{Technical Report} D{\'{e}}partement de
  math{\'{e}}matiques, Institut F{\'{e}}d{\'{e}}ral de Technologie de Zurich,
  Zurich.
\newblock \URLprefix
  \url{http://citeseerx.ist.psu.edu/viewdoc/download?doi=10.1.1.69.792&rep=rep1&type=pdf}
  \bibinfo{note}{accessed on 03-11-2021}.
\bibitem[{Embrechts et~al.(2002)Embrechts, McNeil \& Straumann}]{ref:embrechts}
\bibinfo{author}{Embrechts, P.}, \bibinfo{author}{McNeil, A.}, \&
  \bibinfo{author}{Straumann, D.} (\bibinfo{year}{2002}).
\newblock {\it \bibinfo{title}{{Correlation and dependence in risk management:
  properties and pitfalls}}\/} volume~\bibinfo{volume}{1}.
\newblock \bibinfo{publisher}{Cambridge University Press}.
\bibitem[{{European Banking Authority}(2018)}]{EBA2018}
\bibinfo{author}{{European Banking Authority}} (\bibinfo{year}{2018}).
\newblock \bibinfo{title}{{Sustainable Finance}}.
\newblock \URLprefix
  \url{https://www.eba.europa.eu/financial-innovation-and-fintech/
  sustainable-finance} \bibinfo{note}{accessed on 03-11-2021}.
\bibitem[{{European Banking Authority}(2020)}]{EBADiscussion}
\bibinfo{author}{{European Banking Authority}} (\bibinfo{year}{2020}).
\newblock {\it \bibinfo{title}{{EBA Discussion paper on management and
  supervision of ESG risks for credit institutions and investment firms}}\/}.
\newblock \bibinfo{type}{Technical Report} \bibinfo{number}{October}.
\newblock \URLprefix
  \url{https://www.eba.europa.eu/sites/default/documents/files/document_library/Publications/Discussions/2021/Discussion%20Paper%20on%20management%20and%20supervision%20of%20ESG%20risks%20for%20credit%20institutions%20and%20investment%20firms/935496/2020-11-02%20%20ESG%20Discussion%20Paper.pdf}
  \bibinfo{note}{accessed on 03-11-2021}.
\bibitem[{{European Central Bank}(2021)}]{ECB2021}
\bibinfo{author}{{European Central Bank}} (\bibinfo{year}{2021}).
\newblock {\it \bibinfo{title}{{Climate-related risk and financial
  stability}}\/}.
\newblock \bibinfo{type}{Technical Report}.
\newblock \URLprefix \url{https://www.ecb.europa.eu/pub/pdf/other/
  ecb.climateriskfinancialstability202107{~}87822fae81.en.pdf}
  \bibinfo{note}{accessed on 03-11-2021}.
\bibitem[{Fenech et~al.(2015)Fenech, Vosgha \& Shafik}]{Fenech2015}
\bibinfo{author}{Fenech, J.~P.}, \bibinfo{author}{Vosgha, H.}, \&
  \bibinfo{author}{Shafik, S.} (\bibinfo{year}{2015}).
\newblock \bibinfo{title}{{Loan default correlation using an Archimedean copula
  approach: A case for recalibration}}.
\newblock {\it \bibinfo{journal}{Economic Modelling}\/},  {\it
  \bibinfo{volume}{47}\/}, \bibinfo{pages}{340--354}.
\bibitem[{Fink et~al.(2017)Fink, Klimova, Czado \& St{\"{o}}ber}]{Fink2017}
\bibinfo{author}{Fink, H.}, \bibinfo{author}{Klimova, Y.},
  \bibinfo{author}{Czado, C.}, \& \bibinfo{author}{St{\"{o}}ber, J.}
  (\bibinfo{year}{2017}).
\newblock \bibinfo{title}{{Regime Switching Vine Copula Models for Global
  Equity and Volatility Indices}}.
\newblock {\it \bibinfo{journal}{Econometrics}\/},  {\it
  \bibinfo{volume}{5}\/}, \bibinfo{pages}{1--38}.
\bibitem[{Flori et~al.(2021)Flori, Lillo, Pammolli \& Spelta}]{Flori2019}
\bibinfo{author}{Flori, A.}, \bibinfo{author}{Lillo, F.},
  \bibinfo{author}{Pammolli, F.}, \& \bibinfo{author}{Spelta, A.}
  (\bibinfo{year}{2021}).
\newblock \bibinfo{title}{{Better to stay apart: asset commonality, bipartite
  network centrality, and investment strategies}}.
\newblock {\it \bibinfo{journal}{Annals of Operations Research}\/},  {\it
  \bibinfo{volume}{299}\/}, \bibinfo{pages}{177--213}.
\bibitem[{Frahm et~al.(2005)Frahm, Junker \& Schmidt}]{ref:frahm}
\bibinfo{author}{Frahm, G.}, \bibinfo{author}{Junker, M.}, \&
  \bibinfo{author}{Schmidt, R.} (\bibinfo{year}{2005}).
\newblock \bibinfo{title}{{Estimating the tail-dependence coefficient:
  properties and pitfalls}}.
\newblock {\it \bibinfo{journal}{Insurance: Mathematics and Economics}\/},
  {\it \bibinfo{volume}{37}\/}, \bibinfo{pages}{80--100}.
\bibitem[{Friede(2019)}]{ref:friede}
\bibinfo{author}{Friede, G.} (\bibinfo{year}{2019}).
\newblock \bibinfo{title}{{Why don't we see more action? A metasynthesis of the
  investor impediments to integrate environmental, social, and governance
  factors}}.
\newblock {\it \bibinfo{journal}{Business Strategy and the Environment}\/},
  {\it \bibinfo{volume}{28}\/}, \bibinfo{pages}{1260--1282}.
\bibitem[{Gibson et~al.(2020)Gibson, Krueger \& Schmidt}]{Gibson2020}
\bibinfo{author}{Gibson, R.}, \bibinfo{author}{Krueger, P.}, \&
  \bibinfo{author}{Schmidt, P.~S.} (\bibinfo{year}{2020}).
\newblock \bibinfo{title}{{ESG rating disagreement and stock returns}}.
\newblock {\it \bibinfo{journal}{Swiss Finance Institute Research Paper}\/}, .
  \DOIprefix\doi{10.2139/ssrn.3433728}.
\newblock \bibinfo{note}{Accessed on 03-11-2021}.
\bibitem[{Giese et~al.(2019)Giese, Lee, Melas, Nagy \& Nishikawa}]{ref:giese}
\bibinfo{author}{Giese, G.}, \bibinfo{author}{Lee, L.-E.},
  \bibinfo{author}{Melas, D.}, \bibinfo{author}{Nagy, Z.}, \&
  \bibinfo{author}{Nishikawa, L.} (\bibinfo{year}{2019}).
\newblock \bibinfo{title}{{Foundations of ESG investing: how ESG affects equity
  valuation, risk, and performance}}.
\newblock {\it \bibinfo{journal}{Journal of Portfolio Management}\/},  {\it
  \bibinfo{volume}{45}\/}, \bibinfo{pages}{69--83}.
\bibitem[{{Global Sustainable Investment Alliance (GSIA)}(2018)}]{GSIA2018}
\bibinfo{author}{{Global Sustainable Investment Alliance (GSIA)}}
  (\bibinfo{year}{2018}).
\newblock {\it \bibinfo{title}{{2018 Global Sustainable Investment Review}}\/}.
\newblock \bibinfo{type}{Technical Report}.
\newblock \URLprefix
  \url{http://www.gsi-alliance.org/wp-content/uploads/2019/06/
  GSIR{\_}Review2018F.pdf} \bibinfo{note}{accessed on 03-11-2021}.
\bibitem[{Goldreyer \& Diltz(1999)}]{ref:goldreyer}
\bibinfo{author}{Goldreyer, E.~F.}, \& \bibinfo{author}{Diltz, J.~D.}
  (\bibinfo{year}{1999}).
\newblock \bibinfo{title}{{The performance of socially responsible mutual
  funds: Incorporating sociopolitical information in portfolio selection}}.
\newblock {\it \bibinfo{journal}{Managerial Finance}\/},  {\it
  \bibinfo{volume}{25}\/}, \bibinfo{pages}{23--36}.
\bibitem[{Gormley \& Matsa(2011)}]{ref:gormley2}
\bibinfo{author}{Gormley, T.~A.}, \& \bibinfo{author}{Matsa, D.~A.}
  (\bibinfo{year}{2011}).
\newblock \bibinfo{title}{{Growing out of trouble? Corporate responses to
  liability risk}}.
\newblock {\it \bibinfo{journal}{Review of Financial Studies}\/},  {\it
  \bibinfo{volume}{24}\/}, \bibinfo{pages}{2781--2821}.
\bibitem[{Gormley et~al.(2013)Gormley, Matsa \& Milbourn}]{ref:gormley1}
\bibinfo{author}{Gormley, T.~A.}, \bibinfo{author}{Matsa, D.~A.}, \&
  \bibinfo{author}{Milbourn, T.} (\bibinfo{year}{2013}).
\newblock \bibinfo{title}{{CEO compensation and corporate risk: Evidence from a
  natural experiment}}.
\newblock {\it \bibinfo{journal}{Journal of Accounting and Economics}\/},  {\it
  \bibinfo{volume}{56}\/}, \bibinfo{pages}{79--101}.
\bibitem[{Haff(2012)}]{Haff2012}
\bibinfo{author}{Haff, I.~H.} (\bibinfo{year}{2012}).
\newblock \bibinfo{title}{{Comparison of estimators for pair-copula
  constructions}}.
\newblock {\it \bibinfo{journal}{Journal of Multivariate Analysis}\/},  {\it
  \bibinfo{volume}{110}\/}, \bibinfo{pages}{91--105}.
\bibitem[{Henriksson et~al.(2019)Henriksson, Livnat, Pfeifer \&
  Stumpp}]{ref:henriksson}
\bibinfo{author}{Henriksson, R.}, \bibinfo{author}{Livnat, J.},
  \bibinfo{author}{Pfeifer, P.}, \& \bibinfo{author}{Stumpp, M.}
  (\bibinfo{year}{2019}).
\newblock \bibinfo{title}{{Integrating ESG in portfolio construction}}.
\newblock {\it \bibinfo{journal}{Journal of Portfolio Management}\/},  {\it
  \bibinfo{volume}{45}\/}, \bibinfo{pages}{67--81}.
\bibitem[{Hoepner et~al.(2016)Hoepner, Oikonomou, Sautner, Starks \&
  Zhou}]{ref:hoepner}
\bibinfo{author}{Hoepner, A. G.~F.}, \bibinfo{author}{Oikonomou, I.},
  \bibinfo{author}{Sautner, Z.}, \bibinfo{author}{Starks, L.~T.}, \&
  \bibinfo{author}{Zhou, X.} (\bibinfo{year}{2016}).
\newblock \bibinfo{title}{{{\{}ESG{\}} Shareholder Engagement and Downside
  Risk}}.
\newblock {\it \bibinfo{journal}{SSRN Electronic Journal}\/}, .
  \DOIprefix\doi{10.2139/ssrn.2874252}.
\bibitem[{Joe(1996)}]{ref:joe2}
\bibinfo{author}{Joe, H.} (\bibinfo{year}{1996}).
\newblock \bibinfo{title}{{Families of m-variate distributions with given
  margins and m (m-1)/2 bivariate dependence parameters}}.
\newblock {\it \bibinfo{journal}{Lecture Notes-Monograph Series}\/},  {\it
  \bibinfo{volume}{28}\/}, \bibinfo{pages}{120--141}.
\bibitem[{Joe(2005)}]{ref:joe1}
\bibinfo{author}{Joe, H.} (\bibinfo{year}{2005}).
\newblock \bibinfo{title}{{Asymptotic efficiency of the two-stage estimation
  method for copula-based models}}.
\newblock {\it \bibinfo{journal}{Journal of Multivariate Analysis}\/},  {\it
  \bibinfo{volume}{94}\/}, \bibinfo{pages}{401--419}.
\bibitem[{Joe(2014)}]{ref:joe2014}
\bibinfo{author}{Joe, H.} (\bibinfo{year}{2014}).
\newblock {\it \bibinfo{title}{{Dependence modeling with copulas}}\/}.
\newblock \bibinfo{address}{USA}: \bibinfo{publisher}{CRC press}.
\bibitem[{Joe \& Xu(1996)}]{JoeIFM}
\bibinfo{author}{Joe, H.}, \& \bibinfo{author}{Xu, J.~J.}
  (\bibinfo{year}{1996}).
\newblock {\it \bibinfo{title}{{The Estimation Method of Inference Functions
  for Margins for Multivariate Models}}\/}.
\newblock \bibinfo{type}{Technical Report} \bibinfo{number}{166} Department of
  Statistics, University of British Columbia.
\newblock \DOIprefix\doi{10.14288/1.0225985} \bibinfo{note}{accessed on
  03-11-2021}.
\bibitem[{Jondeau \& Rockinger(2003)}]{ref:jondeau}
\bibinfo{author}{Jondeau, E.}, \& \bibinfo{author}{Rockinger, M.}
  (\bibinfo{year}{2003}).
\newblock \bibinfo{title}{{Testing for differences in the tails of stock-market
  returns}}.
\newblock {\it \bibinfo{journal}{Journal of Empirical Finance}\/},  {\it
  \bibinfo{volume}{10}\/}, \bibinfo{pages}{559--581}.
\bibitem[{Kelly \& Jiang(2014)}]{ref:kelly}
\bibinfo{author}{Kelly, B.}, \& \bibinfo{author}{Jiang, H.}
  (\bibinfo{year}{2014}).
\newblock \bibinfo{title}{{Tail risk and asset prices}}.
\newblock {\it \bibinfo{journal}{Review of Financial Studies}\/},  {\it
  \bibinfo{volume}{27}\/}, \bibinfo{pages}{2841--2871}.
\bibitem[{Kim et~al.(2014)Kim, Li \& Li}]{ref:kim}
\bibinfo{author}{Kim, Y.}, \bibinfo{author}{Li, H.}, \& \bibinfo{author}{Li,
  S.} (\bibinfo{year}{2014}).
\newblock \bibinfo{title}{{Corporate social responsibility and stock price
  crash risk}}.
\newblock {\it \bibinfo{journal}{Journal of Banking {\&} Finance}\/},  {\it
  \bibinfo{volume}{43}\/}, \bibinfo{pages}{1--13}.
\bibitem[{Kurowicka \& Cooke(2006)}]{ref:kura}
\bibinfo{author}{Kurowicka, D.}, \& \bibinfo{author}{Cooke, R.~M.}
  (\bibinfo{year}{2006}).
\newblock {\it \bibinfo{title}{{Uncertainty analysis with high dimensional
  dependence modelling}}\/}.
\newblock \bibinfo{address}{UK}: \bibinfo{publisher}{John Wiley {\&} Sons}.
\bibitem[{Kurowicka \& Joe(2011)}]{ref:kurojoe}
\bibinfo{author}{Kurowicka, D.}, \& \bibinfo{author}{Joe, H.}
  (\bibinfo{year}{2011}).
\newblock {\it \bibinfo{title}{{Dependence modeling - handbook on vine
  copulae}}\/}.
\newblock \bibinfo{address}{Singapore}: \bibinfo{publisher}{World Scientific
  Publishing Co.}
\bibitem[{Li(2000)}]{ref:1}
\bibinfo{author}{Li, D.~X.} (\bibinfo{year}{2000}).
\newblock \bibinfo{title}{{On default correlation: A copula function
  approach}}.
\newblock {\it \bibinfo{journal}{Journal of Fixed Income}\/},  {\it
  \bibinfo{volume}{9}\/}, \bibinfo{pages}{43--54}.
\bibitem[{Li \& Polychronopoulos(2020)}]{Li2020}
\bibinfo{author}{Li, F.}, \& \bibinfo{author}{Polychronopoulos, A.}
  (\bibinfo{year}{2020}).
\newblock \bibinfo{title}{What a difference an esg ratings provider makes}.
\newblock \URLprefix \url{https://www. researchaffiliates.
  com/documents/770-what-a-difference-an-esg-ratings-provider-makes. pdf}
  \bibinfo{note}{accessed on 03-11-2021}.
\bibitem[{Li et~al.(2017)Li, Wang \& Wang}]{ref:li}
\bibinfo{author}{Li, X.}, \bibinfo{author}{Wang, S.~S.}, \&
  \bibinfo{author}{Wang, X.} (\bibinfo{year}{2017}).
\newblock \bibinfo{title}{{Trust and stock price crash risk: Evidence from
  China}}.
\newblock {\it \bibinfo{journal}{Journal of Banking {\&} Finance}\/},  {\it
  \bibinfo{volume}{76}\/}, \bibinfo{pages}{74--91}.
\bibitem[{L{\"{o}}{\"{o}}f et~al.(2021)L{\"{o}}{\"{o}}f, Sahamkhadam \&
  Stephan}]{new:loof2021}
\bibinfo{author}{L{\"{o}}{\"{o}}f, H.}, \bibinfo{author}{Sahamkhadam, M.}, \&
  \bibinfo{author}{Stephan, A.} (\bibinfo{year}{2021}).
\newblock \bibinfo{title}{{Is Corporate Social Responsibility investing a free
  lunch? The relationship between ESG, tail risk, and upside potential of
  stocks before and during the COVID-19 crisis}}.
\newblock \URLprefix \url{http://www.cesis.se} \bibinfo{note}{accessed on
  03-11-2021}.
\bibitem[{Luo \& Bhattacharya(2009)}]{Luo2009}
\bibinfo{author}{Luo, X.}, \& \bibinfo{author}{Bhattacharya, C.~B.}
  (\bibinfo{year}{2009}).
\newblock \bibinfo{title}{{The debate over doing good: Corporate social
  performance, strategic marketing levers, and firm-idiosyncratic risk}}.
\newblock {\it \bibinfo{journal}{Journal of Marketing}\/},  {\it
  \bibinfo{volume}{73}\/}, \bibinfo{pages}{198--213}.
\bibitem[{Maiti(2020)}]{Maiti2020}
\bibinfo{author}{Maiti, M.} (\bibinfo{year}{2020}).
\newblock \bibinfo{title}{{Is ESG the succeeding risk factor?}}
\newblock {\it \bibinfo{journal}{Journal of Sustainable Finance {\&}
  Investment}\/},  {\it \bibinfo{volume}{11}\/}, \bibinfo{pages}{1--15}.
\bibitem[{Malevergne et~al.(2005)Malevergne, Pisarenko \&
  Sornette}]{ref:malevergne}
\bibinfo{author}{Malevergne, Y.}, \bibinfo{author}{Pisarenko, V.}, \&
  \bibinfo{author}{Sornette, D.} (\bibinfo{year}{2005}).
\newblock \bibinfo{title}{{Empirical distributions of stock returns: between
  the stretched exponential and the power law?}}
\newblock {\it \bibinfo{journal}{Quantitative Finance}\/},  {\it
  \bibinfo{volume}{5}\/}, \bibinfo{pages}{379--401}.
\bibitem[{Minor(2011)}]{ref:minor}
\bibinfo{author}{Minor, D.~B.} (\bibinfo{year}{2011}).
\newblock \bibinfo{title}{{Corporate citizenship as insurance: Theory and
  evidence}}.
\newblock {\it \bibinfo{journal}{University of California, Berkley}\/}, .
\bibitem[{Nagler et~al.(2019)Nagler, Bumann \& Czado}]{Nagler2019}
\bibinfo{author}{Nagler, T.}, \bibinfo{author}{Bumann, C.}, \&
  \bibinfo{author}{Czado, C.} (\bibinfo{year}{2019}).
\newblock \bibinfo{title}{{Model selection in sparse high-dimensional vine
  copula models with an application to portfolio risk}}.
\newblock {\it \bibinfo{journal}{Journal of Multivariate Analysis}\/},  {\it
  \bibinfo{volume}{172}\/}, \bibinfo{pages}{180--192}.
\bibitem[{Naifar(2012)}]{Naifar2012}
\bibinfo{author}{Naifar, N.} (\bibinfo{year}{2012}).
\newblock \bibinfo{title}{{Modeling the dependence structure between default
  risk premium, equity return volatility and the jump risk: Evidence from a
  financial crisis}}.
\newblock {\it \bibinfo{journal}{Economic modelling}\/},  {\it
  \bibinfo{volume}{29}\/}, \bibinfo{pages}{119--131}.
\bibitem[{Nguyen \& Bhatti(2012)}]{Nguyen2012}
\bibinfo{author}{Nguyen, C.~C.}, \& \bibinfo{author}{Bhatti, M.~I.}
  (\bibinfo{year}{2012}).
\newblock \bibinfo{title}{{Copula model dependency between oil prices and stock
  markets: Evidence from China and Vietnam}}.
\newblock {\it \bibinfo{journal}{Journal of International Financial Markets,
  Institutions and Money}\/},  {\it \bibinfo{volume}{22}\/},
  \bibinfo{pages}{758--773}.
\bibitem[{Nofsinger \& Varma(2014)}]{ref:nofsinger}
\bibinfo{author}{Nofsinger, J.}, \& \bibinfo{author}{Varma, A.}
  (\bibinfo{year}{2014}).
\newblock \bibinfo{title}{{Socially responsible funds and market crises}}.
\newblock {\it \bibinfo{journal}{Journal of Banking {\&} Finance}\/},  {\it
  \bibinfo{volume}{48}\/}, \bibinfo{pages}{180--193}.
\bibitem[{Pourkhanali et~al.(2016)Pourkhanali, Kim, Tafakori \&
  Fard}]{Pourkhanali2016}
\bibinfo{author}{Pourkhanali, A.}, \bibinfo{author}{Kim, J.-M.},
  \bibinfo{author}{Tafakori, L.}, \& \bibinfo{author}{Fard, F.~A.}
  (\bibinfo{year}{2016}).
\newblock \bibinfo{title}{{Measuring systemic risk using vine-copula}}.
\newblock {\it \bibinfo{journal}{Economic modelling}\/},  {\it
  \bibinfo{volume}{53}\/}, \bibinfo{pages}{63--74}.
\bibitem[{Puccetti \& Scherer(2018)}]{ref:puc}
\bibinfo{author}{Puccetti, G.}, \& \bibinfo{author}{Scherer, M.}
  (\bibinfo{year}{2018}).
\newblock \bibinfo{title}{{Copulas, credit portfolios, and the broken heart
  syndrome}}.
\newblock {\it \bibinfo{journal}{Dependence Modeling}\/},  {\it
  \bibinfo{volume}{6}\/}, \bibinfo{pages}{114--130}.
\bibitem[{Refinitiv(2021)}]{Refinitiv2021}
\bibinfo{author}{Refinitiv} (\bibinfo{year}{2021}).
\newblock \bibinfo{title}{{Environmental, Social and Governance (ESG) scores
  from Refinitiv}}.
\newblock \URLprefix
  \url{https://www.refinitiv.com/content/dam/marketing/en_us/documents/methodology/refinitiv-esg-scores-methodology.pdf}
  \bibinfo{note}{accessed on 03-11-2021}.
\bibitem[{Sahin et~al.(2021)Sahin, Bax, Paterlini \& Czado}]{Sahin2021}
\bibinfo{author}{Sahin, {\"{O}}.}, \bibinfo{author}{Bax, K.},
  \bibinfo{author}{Paterlini, S.}, \& \bibinfo{author}{Czado, C.}
  (\bibinfo{year}{2021}).
\newblock \bibinfo{title}{{ESGM: ESG scores and the Missing pillar}}.
\newblock {\it \bibinfo{journal}{SSRN Electronic Journal}\/}, .
  \DOIprefix\doi{10.2139/ssrn.3890696}.
\bibitem[{Salmon(2012)}]{ref:2}
\bibinfo{author}{Salmon, F.} (\bibinfo{year}{2012}).
\newblock \bibinfo{title}{{The formula that killed Wall Street}}.
\newblock {\it \bibinfo{journal}{Significance}\/},  {\it
  \bibinfo{volume}{9}\/}, \bibinfo{pages}{16--20}.
\bibitem[{Schwarz(1978)}]{Schwarz1978}
\bibinfo{author}{Schwarz, G.} (\bibinfo{year}{1978}).
\newblock \bibinfo{title}{{Estimating the dimension of a model}}.
\newblock {\it \bibinfo{journal}{The Annals of Statistics}\/},  {\it
  \bibinfo{volume}{6}\/}, \bibinfo{pages}{461--464}.
\bibitem[{Serafeim \& Yoon(in press)}]{Serafeim2021b}
\bibinfo{author}{Serafeim, G.}, \& \bibinfo{author}{Yoon, A.}
  (\bibinfo{year}{in press}).
\newblock \bibinfo{title}{{Stock Price Reactions to ESG News: The Role of ESG
  Ratings and Disagreement}}.
\newblock {\it \bibinfo{journal}{Review of Accounting Studies}\/}, .
  \DOIprefix\doi{10.2139/ssrn.3765217}.
\newblock \bibinfo{note}{Accessed on 03-11-2021}.
\bibitem[{Shafer \& Szado(2018)}]{ref:shafer}
\bibinfo{author}{Shafer, M.}, \& \bibinfo{author}{Szado, E.}
  (\bibinfo{year}{2018}).
\newblock \bibinfo{title}{{Environmental, social, and governance practices and
  perceived tail risk}}.
\newblock {\it \bibinfo{journal}{Accounting {\&} Finance}\/},  {\it
  \bibinfo{volume}{60}\/}, \bibinfo{pages}{4195--4224}.
  \DOIprefix\doi{10.1111/acfi.12541}.
\bibitem[{Sherwood \& Pollard(2017)}]{Sherwood2017}
\bibinfo{author}{Sherwood, M.~W.}, \& \bibinfo{author}{Pollard, J.~L.}
  (\bibinfo{year}{2017}).
\newblock \bibinfo{title}{{The risk-adjusted return potential of integrating
  ESG strategies into emerging market equities}}.
\newblock {\it \bibinfo{journal}{Journal of Sustainable Finance {\&}
  Investment}\/},  {\it \bibinfo{volume}{8}\/}, \bibinfo{pages}{26--44}.
\bibitem[{Shirvani \& Volchenkov(2019)}]{ref:shir}
\bibinfo{author}{Shirvani, A.}, \& \bibinfo{author}{Volchenkov, D.}
  (\bibinfo{year}{2019}).
\newblock \bibinfo{title}{{A regulated market under sanctions: On tail
  dependence between oil, gold, and Tehran stock exchange index}}.
\newblock \URLprefix \url{https://arxiv.org/abs/1911.01826}.
  \DOIprefix\doi{arXiv:1911.01826} \bibinfo{note}{accessed on 03-11-2021}.
\bibitem[{Sklar(1959)}]{ref:sklar}
\bibinfo{author}{Sklar, A.} (\bibinfo{year}{1959}).
\newblock \bibinfo{title}{{Fonctions de repartition an dimensions et leurs
  marges}}.
\newblock {\it \bibinfo{journal}{Publications de l'Institut de Statistique de
  l'Universite de Paris}\/},  {\it \bibinfo{volume}{8}\/},
  \bibinfo{pages}{229--231}.
\bibitem[{Wamba et~al.(2020)Wamba, Sahut, Braune, Teulon \& Others}]{ref:wamba}
\bibinfo{author}{Wamba, L.~D.}, \bibinfo{author}{Sahut, J.-M.},
  \bibinfo{author}{Braune, E.}, \bibinfo{author}{Teulon, F.}, \&
  \bibinfo{author}{Others} (\bibinfo{year}{2020}).
\newblock \bibinfo{title}{{Does the optimization of a company's environmental
  performance reduce its systematic risk? New evidence from European listed
  companies}}.
\newblock {\it \bibinfo{journal}{Corporate Social Responsibility and
  Environmental Management}\/},  {\it \bibinfo{volume}{27}\/},
  \bibinfo{pages}{1677--1694}.
\bibitem[{Xu \& Li(2009)}]{ref:xu}
\bibinfo{author}{Xu, Q.}, \& \bibinfo{author}{Li, X.-M.}
  (\bibinfo{year}{2009}).
\newblock \bibinfo{title}{{Estimation of dynamic asymmetric tail dependences:
  an empirical study on Asian developed futures markets}}.
\newblock {\it \bibinfo{journal}{Applied Financial Economics}\/},  {\it
  \bibinfo{volume}{19}\/}, \bibinfo{pages}{273--290}.
\bibitem[{Zhang et~al.(2021)Zhang, {De Spiegeleer} \& Schoutens}]{Zhang2021}
\bibinfo{author}{Zhang, J.}, \bibinfo{author}{{De Spiegeleer}, J.}, \&
  \bibinfo{author}{Schoutens, W.} (\bibinfo{year}{2021}).
\newblock \bibinfo{title}{{Implied Tail Risk and ESG Ratings}}.
\newblock {\it \bibinfo{journal}{Mathematics}\/},  {\it \bibinfo{volume}{9}\/},
  \bibinfo{pages}{1611}.

\end{thebibliography}

\end{document}